\documentclass[useAMS,usenatbib]{mn2e}

\usepackage{graphicx}

\usepackage{url,times,stfloats,amsmath,amsfonts,amssymb,aas_macros,color,epsfig,epstopdf}
\usepackage[colorlinks=true, citecolor=blue, linkcolor=blue, urlcolor=blue, pdfborder={0 0 0}]{hyperref}

\usepackage{booktabs}
\usepackage{multirow}
\usepackage{float}
\usepackage{rotating}
\usepackage{soul}

\newcommand{\LCDM}{$\Lambda$CDM}

\newcommand{\hMsun}{{\ifmmode{h^{-1}{\rm {M_{\odot}}}}\else{$h^{-1}{\rm{M_{\odot}}}$}\fi}}
\newcommand{\lsim}{\lower.5ex\hbox{\ltsima}}
\newcommand{\gsim}{\lower.5ex\hbox{\gtsima}}

\newcommand{\Tab}[1]{Table~\ref{#1}}
\newcommand{\Sec}[1]{Section~\ref{#1}}

\newcommand{\Fig}[1]{Fig.~\ref{#1}}

\def\head{
 \vbox to 0pt{\vss
                   \hbox to 0pt{\hskip 440pt\rm LA-UR-10-07069\hss}
                  \vskip 25pt}}

\title[Planes of satellites in the MW and M31]
{An updated detailed characterization of planes of satellites in the MW and M31}
\author[I. Santos-Santos et al.]
       {Isabel M. Santos-Santos$^{1,2}$\thanks{E-mail: isantos@uvic.ca},           
       Rosa Dom\'inguez-Tenreiro$^{2,3}$,
       Marcel S. Pawlowski$^{4}$\\
$^{1}$Department of Physics and Astronomy, University of Victoria, Victoria, BC, Canada V8P 5C2\\
$^{2}$Departamento de F\'isica Te\'orica, Universidad Aut\'onoma de Madrid, 28049 Cantoblanco, Madrid, Spain\\
$^{3}$Centro para la Investigación Avanzada en Física Fundamental, CIAFF-UAM; Universidad Aut\'onoma de Madrid, 28049 Cantoblanco, Madrid, Spain\\
$^{4}$Leibniz-Institute f\"ur Astrophysik Potsdam, An der Sternwarte 16, D-14482 Potsdam, Germany
}

\begin{document}

\date{Accepted XXXX . Received XXXX; in original form XXXX}

\pagerange{\pageref{firstpage}--\pageref{lastpage}} \pubyear{2018}

\maketitle

\label{firstpage}

%\clearpage

%%%%%%%%%%%%%%%%%%%%%%%%%%%%%%%%%%%%%%%%%%%%%%%%%%%
\begin{abstract}

We present a detailed characterization of planes of satellite galaxies in the Milky Way (MW) and M31. For a positional analysis, we introduce an extension to the `4-galaxy-normal density plot' method \citep[][P13]{Pawlowski13}. It finds the normal directions to the predominant planar configurations of satellites of a system, yielding for each a \textit{collection} of planes of increasing member satellites. 
This allows to quantify the quality of planes in terms of population ($N_{\rm sat}$) and spatial flattening ($c/a$).  
 We apply this method to the latest data for confirmed MW and M31 satellite samples, with 46 and 34 satellites, respectively.
New MW satellites form part of planes previously identified from the sample with  $N_{\rm sat}=27$ studied in P13:
 we identify a new plane with $N_{\rm sat}=39$ as thin as the VPOS-3 ($c/a\sim 0.2$), and with roughly the same normal direction; so far the most populated plane that thin reported in the Local Group. 
We introduce a new method to determine, using kinematic data, the axis of maximum co-orbitation of MW satellites. Interestingly, this axis approximately coincides with the normal to the former plane: $\geq45\pm5\%$ of satellites co-orbit. 
In M31 we discover a plane with $N_{\rm sat}=18$ and $c/a\sim0.15$, i.e., quality comparable to the GPoA, and perpendicular to it. 
This structure is viewed face-on from the Sun making it susceptible to M31 satellite distance uncertainties. An estimation of the perpendicular velocity dispersion suggests it is dynamically unstable. Finally we find that mass is not a property determining a satellite's  membership to good quality planes.

\end{abstract}

%%%%%%%%%%%%%%%%%%%%%%%%%%%%%%%%%%%%%%%%%%%%%%%%%%%
\noindent
\begin{keywords}
galaxies: dwarf - Local Group - kinematics and dynamics  methods: statistical cosmology: theory
 \end{keywords}

%%%%%%%%%%%%%%%%%%%%%%%%%%%%%%%%%%%%%%%%%%%%%%%%%%%

\section{Introduction}
\label{Intro}

The known objects surrounding the Milky Way (MW) show an anisotropic spatial distribution.
\citet{Lynden76}
and \citet{Kunkel76} were
  the first to notice that 
dSphs Sculptor, Draco, Ursa Minor  and globular cluster Palomar 13, apparently lied on the orbital plane of  the Magellanic stream, therefore polar to the Galaxy.
Soon after, Fornax, Leo I, Leo II, Sextans, Phoenix
and some classified as `young halo' globular clusters, 
 were   also found to participate of this ``great circle" \citep{LyndenBell82,Majewski94,FusiPecci95}.
 The existence of a ``plane of satellites" 
in the MW
 was finally ratified when measuring the 
  very flattened distribution of the  11 classical  satellites as compared to isotropy \citep{Kroupa05,Metz07}.
In our neighboring galactic system Andromeda (M31), 
the first studies 
  by \citet{Grebel99} and \citet{Koch06}
on the then-known $\lesssim$15 dwarf galaxies within $\sim$500 kpc distance,
found 
that the subsample of dSph/dE type dwarfs 
also lied near a ``great circle". %% \citep{Grebel99,Koch06}. 
This  spatial anisotropy 
was emphasized by the skewness of M31 dwarfs in the direction to the MW
 \citep{McConnachie06}.
More recently,
the addition to the picture
 of newly discovered 
faint
 satellites
 thanks to surveys like SDSS \citep{York00} or PAndAS \citep{McConnachie09},
 and an increased quality of distance measurements, %% for the already known dwarfs,
 has only enhanced the significance 
 of the planar structures noted in the MW and M31 \citep{Metz09,Kroupa10,Ibata13,Conn13,Pawlowski13,Pawlowski14,Pawlowski15b}. 
In addition,
  a richer census of young halo globular clusters and several stellar/gaseous streams 
 have been
  shown to also align with the MW satellites \citep{Keller12,Pawlowski12,Riley2020}. 
Finally, apart from the MW and M31, there are  claims for  a planar distribution of satellites in the nearby Centaurus A Group of galaxies \citep{Tully15}, further supported by discoveries of new dwarfs and better distance estimates \citep{Muller16,Muller18}.

 In the last years,
the quantification of 
  such    planar alignments 
  has gained increasing importance 
 in order to
 unambiguously define the observed structures
in terms of their orientation and characteristics.
These quantifications have  demanded increasingly more sophisticated methods that make use of the three-dimensional position data as well as statistical approaches to overcome measurement uncertainties or avoid spurious effects coming from working with a small sample number.

Specifically,
\citet{Koch06} used an error-weighted orthogonal distance regression accompanied by bootstrapped tests, to reliably determine a robust solution for estimating best-fit planes. 
They fitted a plane to all possible subsamples of M31 satellites 
involving 3 to 15 members
and projected the resulting normal vectors on a sphere. With the density distribution of normals they found an estimation of the normal direction to 
the  broad planar distribution defined by the satellites' positions.
\citet{Metz07,Metz09} used instead the Tensor of Inertia  plane-fitting method
taking into account distance uncertainties. 
More recently, \citet{Pawlowski13},
(hereafter P13),
 combined the previous efforts and  presented a new statistical method to define the direction of   predominant plane-like spatial distributions  of satellites within a given sample:  the `4-galaxy-normal density plot' method, consisting of a planar fit to every combination of 4 satellites. 
 From its application  to the 
 confirmed satellites 
within 300 kpc in the MW and M31
at that time, they found 
 %they obtained the normal direction to 
one predominant planar alignment of satellites in each galactic system 
and measured their respective normal directions. 
In this way, 
the so-called ``VPOS-3" (Vast Polar Structure) in the MW and ``GPoA" (Great Plane of Andromeda) in M31 planes
 were defined, consisting of 24 and 19 satellites respectively.
These planes have been considered so far to be the most relevant satellite planar configurations in the MW and M31, 
and have been used as a benchmark against which to test the
 alignment of the newest (some even of  unclassified nature) objects discovered \citep{Pawlowski14,Pawlowski15b}. 
%%% However, an identification of the ``best" planes of satellites in the MW and M31  formed by a variable number of members is still lacking. 
%\textbf{However, a detailed characterization of these predominant planar structures in the MW and M31, highlighting the ``best" planes of satellites  formed by a variable number of members, is still lacking.}

However, a detailed characterization of these predominant planar structures in the MW and M31, finding which are the combinations of $N_{\rm sat}$ satellites out of a total sample $N_{\rm tot}$ that conform the most relevant planar structures, is still lacking. Moreover, it is of interest to highlight those specific planes of satellites within a system with highly striking characteristics.
This identification demands an analysis of how the \textit{quality} of these planes changes with the number of satellites involved.

This paper is the first of a series where the results of a project aimed at working in depth the plane of satellites issue
within a \LCDM\, cosmological context
 are reported.
The purpose of this paper is to present a more detailed quantification and characterization of the plane-like spatial structures in the MW and M31 satellite systems. 
This is an important issue and,
additionally, it will provide a reference with which to compare results from 
  analyses of numerical simulations of the formation of  disc galaxy systems
\citep[][hereafter Paper II]{SantosSantos2020}.
%%% (Santos-Santos et al. 2020, hereafter Paper II).}
%
We  focus on the positions of satellites, which for the MW have error bars  much smaller than those of kinematic data. 
We note, however, that complementary, relevant information to satellite planar alignments comes from kinematic data.
Indeed, recent proper motions for MW satellites measured with \textit{Gaia} 
\citep[see e.g.,][]{GaiaHelmi18,Fritz18,Simon2018}
  suggest that a non-negligible fraction of them are orbiting along the VPOS plane \citep{Fritz18} 
%However, also a high fraction of them  
while others are most likely not associated to the structure as their orbital poles are not aligned with the VPOS normal vector \citep[see e.g.,][for a recent study of the classical MW satellites]{Pawlowski2020}. 
To study the reproducibility of these results in numerical simulations, in Santos-Santos et al. (in preparation, hereafter Paper III) we develop in depth 
the so-called `3J$_{\rm orb}$-barycenter' method, and apply it to zoom-in simulations of disc  galaxies. This method identifies  the axes around which a maximum number of satellites, out of a given population, co-orbit; allowing the study of relevant issues such  as plane persistence along galaxy evolution.

%%To complete the analyses of positionally-identified planes in the MW, we briefly analyze here their maximum rotation axes and the co-orbitation issue.  
%%For M31 satellites  
%%just line-of-sight velocities \citep{Ibata13} are currently available and not such analyses a possible yet.}

In this paper we build on the results of 
P13
%\citet{Pawlowski13} 
and develop
an extension to the `4-galaxy-normal density plot' method that enables a deeper
% study on 
cataloguing and quality analyses of planes of satellites in a galactic system. 
In particular,
for each predominant planar configuration of satellites 
in the MW and M31
found with the previous method,
we yield a collection of planes 
%of satellites with an increasing number of members,
of increasing $N_{\rm sat}$ satellite members,
and we identify the highest-quality ones
%planes   
in terms of the Tensor of Inertia parameters
and the number of satellites involved.
In particular, a plane's quality is quantified in terms of its population ($N_{\rm sat}$) and flattening (the concentration ellipsoid short-to-long axis ratio $c/a$ or, equivalently, the r.m.s. thickness normal to the plane; \citealt{Cramer}). Quality of planes with the same population or flattening can be compared with each other, allowing to single-out high quality planes with the same $N_{\rm sat}$ or $c/a$ values.

Finally, to complete the analyses of positionally-identified planes of satellites in the MW, we briefly study here their
%kinematical-coherence
kinematically-coherent
 character, and independently, identify the
 MW's 
 maximum %rotation
  satellite co-orbitation axis 
%and the co-orbitation issue 
through the `3J$_{\rm orb}$-barycenter' method mentioned above.
For M31 satellites only line-of-sight velocities are currently available \citep{Ibata13}  and such a study is not  possible yet. In addition, distances to M31 satellites remain highly uncertain \citep[see e.g.][]{Weisz2019}.

 The paper is organized as follows. 
In Section~\ref{observedplanes}  we present the samples and datasets of MW and M31 satellites studied.
 In Section~\ref{4gnd_method} we thoroughly describe 
our methodology for plane-searching from satellite position data. 
 Sections \ref{ResultsMW} and \ref{M31}  show the results of our quality analysis for planes in the MW and M31, respectively.
The co-orbitation issue and the determination of the axes of maximum rotation in the MW are addressed in Section~\ref{sec6D}.  Finally, Section~\ref{conclu} summarizes our conclusions.

\section{MW and M31 data}\label{observedplanes}

%\textbf{Consistently with the aim of this paper (within the series reporting  on results of our project, see Section~\ref{Intro} for further details)}
In this project, 
we analyze
planes of satellites in
 samples of $N_{\rm tot}$ confirmed \textit{satellite} (i.e., bound to the MW or M31 hosts) \textit{galaxies} (i.e., not other objects such as globular clusters or streams have been considered). 

For the MW we use two different satellite samples. First, for the sake of consistency with P13,
%\citet{Pawlowski13}, 
and for simplicity as far as comparisons are concerned \citep[see e.g.][]{Pawlowski16}, we analyze the same satellite sample as in 
%the former paper, 
P13 consisting of $N_{\rm tot} = 27$ satellites. This satellite sample will be hereafter termed 'MW27 sample' and it
consists of the classical and SDSS satellites.
% considers all the classical plus SDSS satellites. 
MW27 lists all the satellites  within 300 kpc from the MW, according to  
the   \citet{McConnachie12} ``Nearby dwarf galaxy database"\footnote{\url{http://www.astro.uvic.ca/~alan/Nearby_Dwarf_Database_files/NearbyGalaxies.dat}}
 %%%\citet{McConnachie12}'s database  as  
 as of June 2013. 
   Most probable satellite position values 
and their corresponding Gaussian width uncertainties in the radial Sun - satellite distance
have been taken from
this database,
 as summarized in table 2 of P13.
 %\citet{Pawlowski13}.
 Canis Major   and BootesIII
%%%and AXXVII 
 are considered in MW27 as dwarf galaxies although their nature as such   is
ruled out \citep[see][]{Momany04,MartinezDelgado05,Mateu09},
and   debated \citep[see][]{Grillmair2009,Carlin2018},
respectively.

On top of this historical sample, we also analyze 
an up-to-date (as of August 2020) sample of all confirmed MW satellite galaxies,   consisting of   $N_{\rm tot} = 46$
satellites; hereafter 'MW46'.
The MW46 sample includes 25 satellites in common with MW27
(Canis Major and BootesIII are not included as they are  of doubtful nature, see above), 
and 21 new ones. 
These are recently discovered ultrafaint  satellite galaxies within a radial distance of 300 kpc from the MW center
which originate from a variety of sources/surveys 
 \citep[e.g. The Dark Energy Survey, DECam surveys like Maglites, DELVE, ATLAS, Subaru/Hyper Suprime-Cam Survey, \textit{Gaia}][]{DES15,Koposov15a,Drlica-Wagner2015,Homma2016,Torrealba2016,Torrealba2019,Mau2020}.
In particular, here we consider as "confirmed" ultrafaint MW satellites those objects that are
either
 spectroscopically confirmed \citep[see][]{Simon2019}, 
 or,
alternatively,  
  that populate the dwarf galaxy parameter space of the $M_V$-r$_{\rm half}$ plane (relation between absolute V magnitude and half-light radius of the object), in opposition to the globular cluster regime (note this is only possible to assess in the cases where r$_{\rm half}$ data is available).
 These 21 objects are:
 Tucana2, Hydra2, Hydrus1, Carina2, Pictor2, Virgo1, Aquarius2, Crater2, Antlia2, BootesIV, Cetus3,
HorologiumI, ReticulumII, PegasusIII, Tucana4, Columba1, Grus2, PheonixII, HorologiumII, ReticulumIII, Centaurus1.\footnote{
Of the remaining objects listed in the \citet{McConnachie12} database, we do not consider as  confirmed MW satellite galaxies: CarinaIII , Pictoris1, GrusI, SagittariusII, Tucana5, Draco2, Cetus2, Triangulum2, Eridanus3, Indus1 (as they present a small size for which their nature as a galaxy is doubtful), Tucana3 and HydraI (because of their current state as a very tidally disrupted satellite with a stream, which suggests its original orbital behaviour may have been critically altered), or Indus2 (although large in size, the discovery paper claims it is a low confidence detection). In addition, the objects without measured effective radii are not considered as it is not possible to asses their loci in the $M_V-$r$_{\rm half}$ plane.}
In Appendix \ref{sec:obsdata} we present Table~\ref{tab:mwdata} with the positions (RA, dec),    distances from the Sun and stellar masses\footnote{
Observational stellar masses have been computed applying the \citet{Woo08} mass-to-light ratios 
according to galaxy morphological type,
to the V-band luminosities in \citet{McConnachie12}.}  of all MW satellites used in this paper.

As for M31, 
%for the reasons given above, 
we use as reference the same satellite sample as  in 
P13, consisting of 34  confirmed satellites within 300 kpc of the M31 center \citep[hereafter M31\_34, see][]{McConnachie12}. As of August 2020, no new M31 confirmed satellites within 300 kpc  have been added  to those listed in P13's 
table 2.
For completeness, 
%following some author´s suggestions, 
as the mass of M31 is estimated to be larger than that of the MW, and therefore M31 may present a larger virial radius,
we have also considered in our analysis the case of extending  the limiting distance to 350 kpc. This adds two extra satellites to the sample,   AXVI and AXXXIII (also known as PerseusI), at $\sim323$ and $\sim346$ kpc from M31, respectively. We will refer to this extended sample as M31\_36.

In Table~\ref{tab:m31data} we list the M31 satellites included in our analysis and the data used:
RA and dec positions,
%the heliocentric 
distances from the Sun,
the references the distances assumed come  from (with the specific distance determination method) 
% the method their respective distance determinations are based on and specific distance references, 
and the stellar masses. 
Most heliocentric distances in this Table \citep[as compiled in  the database by ][]{McConnachie12} are based on
ground-based tip of the red giant branch
 (TRGB) calibration; however, in some cases they are not.
 We note that this may introduce a bias, 
as distances for the same object obtained with different methods  have been  shown to sometimes be very different   \citep[compare for example][]{McConnachie05,Conn2012,Weisz2019}.
In order to have results where the distance determinations are based on an homogeneous method, we have repeated the analyses with \citet{Conn2012} data (see their table 2), where distances for a long list of M31 satellites were obtained using the TRGB standard candle.
In this case, for the 11 out of 36  M31 satellites that are not in \citet{Conn2012}, distances in Table~\ref{tab:m31data} have been used.\footnote{\citet{Weisz2019} give \textit{HST}-based HB and TRGB  distances to 17 M31 satellites. Unfortunately, these  are typically 0.1 - 0.2 magnitudes further away than their corresponding  ground-based TRGB determinations, because the calibration of the methods is still an issue (see their fig.2).
This data is thus useless to combine with the  TRGB ground-based distances of the remaining M31 satellites.}
 
 %\footnote{\citet{Weisz2019} give \textit{HST}-based distances to 17 M31 satellites. Unfortunately they  are typically 0.1 - 0.2 magnitudes further away than their corresponding  ground-based TRGB determinations. Indeed, these are relative distances where the zero-point is an issue, making them useless to combine with the ground-based determinations of the remaining M31 satellites.}

The different observational planes of satellites claimed  in the literature 
which we will compare to are listed in Table~\ref{table_obspl}
(see next section for definitions of plane properties). 
These were defined in  
%\citet{Pawlowski13}  
P13
considering the 
MW27 and M31\_34 samples of satellites.
%This Table shows the  properties (see next section for property definitions)  of these observed planes as reported by the  most up-to-date studies.  
%
For the MW these planes are: the so-called `classical' \citep[i.e., the 11 most luminous MW satellites, see][]{Metz07},  
the `VPOSall' 
%\citep[][defined by all the 27 confirmed satellites within 300 kpc]{Pawlowski13}, 
(defined by all the 27 confirmed satellites within 300 kpc)
and the
`VPOS-3' 
%\citep[][defined by 24 out of 27 of the VPOSall satellites]{Pawlowski13}.
(defined by 24 out of 27 of the VPOSall satellites).
For M31, there is the plane of satellites noted by \citet{Ibata13} and \citet{Conn13}  with the PAndAS survey,
 which we will consider with 14 satellites
 % instead of 15 
% \citep[as analyzed in][hereafter the `Ibata-Conn-14' plane]{Pawlowski13}
(as analyzed in P13, hereafter the `Ibata-Conn-14' plane)
 \footnote{\citet{Pawlowski13} did not consider AXVI as an M31 satellite because it is further than 300 kpc away from M31. Note that in this paper we include it in the M31\_36 sample.}, and the so-called `GPoA' 
 %\citep{Pawlowski13},
  with 19 members (the 14 of the `Ibata-Conn-14' plane plus 5 more).\footnote{We note that other planes of satellites in the Local Group have been suggested in \citet{Shaya13}, but under the consideration of a different initial sample of satellites than that used here. In particular, they define 4 satellite planes (2 in the MW and 2 in M31). The so-called ``plane 1" includes a majority of satellites that   participate in the GPoA, while ``plane 4" is basically a reduced version of the classical plane in the MW plus dwarf galaxy Phoenix.}

\section{ Planar configurations from position data}
\label{4gnd_method}

Our method to find planar structures and assess their quality consists of 2 parts. The first part follows the 
'4-galaxy-normal density plot' method described in \citet{Pawlowski13}. 
This technique checks if there is a subsample of a given satellite sample that defines 
a dominant
 planar arrangement 
  in terms of the outputs of 
the standard Tensor of Inertia (ToI) plane-fitting technique   \citep[see][]{{Metz07,Pawlowski13}}, 
based on an orthogonal-distance regression.
In terms of the corresponding concentration ellipsoid \citep{Cramer}, planes
are  characterized by:
\begin{itemize}   
\item  $N_{\rm sat}$: the number of satellites in the subsample;
\item $\vec{n}$, the normal to the best fitting plane; 
\item $c/a$: the ellipsoid short-to-long axis ratio;
\item $b/a$:  the ellipsoid intermediate-to-long axis ratio;
\item $\Delta$RMS: the root-mean-square thickness perpendicular to the best-fitting plane;
\item $D_{\rm cg}$: the distance from the center-of-mass of the main galaxy to the plane.
\end{itemize}
These outputs are used to quantify the quality of planes.
To begin with, a planar configuration must be flattened (i.e., low $c/a$), and, as opposed to filamentary, it also requires $b/a \sim 1$  for an oblate distribution.
High quality planes are those
 with a high $N_{\rm sat}$, and a low $c/a$ and $\Delta$RMS,
meaning they  are populated and thin (plane  quality as understood in this work will be specified in more detail at the end of the next section).
% Moreover,  the plane normal $\vec{n}$ determines the plane direction, for example in view of Aitoff projection purposes.  
 Finally,
  a low $D_{\rm cg}$ means that the plane passes near the main galaxy's center,
a characteristic that dynamically stable planes must show, assuming that the host galaxy's center is close to the center of the system's gravitational potential well.
 % a characteristic to be requested  if the planes are expected to live within a potential making them  dynamically stable, assuming that the host galaxy center is close to the center of the system's gravitational potential well.  

The second part of our methodology, which is the focus of this paper, is an extension to the
4-galaxy-normal density plot 
% previously mentioned 
method, consisting of a \textit{quality} analysis of the predominant planar arrangements found.

\subsection{4-galaxy-normal density (4GND) plot method}\label{N4p}
This method was presented in sec. 2.4 of \citet{Pawlowski13}. We briefly summarize it and mention the procedure  particularities followed in this study.

\begin{enumerate}

\item 
A plane is fitted to every combination of four\footnote{
Three points always define a plane, not allowing any quantification of plane thickness.
Therefore $4$ is the lowest possible amount to take into consideration under the condition of making the number of combinations
high enough to get a good outcome signal. While a larger number could be used, this choice allows 
%as high as possible. This is important in order 
to analyze sets of satellites consisting of a low number of  objects, as  is frequently the case in satellite populations.
}
 satellites' positions, using the ToI  technique. The resultant normal vector (i.e. 4-galaxy-normal) and corresponding plane parameters are stored.
 To account for distance uncertainties, this step is repeated 100 times using 100 random positions per satellite,  calculated
 using their corresponding radial distance uncertainties.

\item
All the 4-galaxy-normals 
(from all 100 realizations)
 are projected on a regularly-binned sphere, assuming a Galactocentric coordinate system such that the MW's disc spin vector points towards the south pole. 
A density map (i.e. 2D-histogram)  is drawn from the projections, where 
each normal has been weighted by $\log\left( \frac{a+b}{c} \right)$ to emphasize planar-like spatial distributions.
The over-density regions   in these density maps  (i.e. regions of  4-galaxy-normal accumulation) 
therefore signal the normal direction to 
a dominant planar space-configuration. Satellites contributing 4-galaxy-normals to a given over-density are likely members of such a dominant plane. 
As opposite normal vectors indicate the same plane, density maps  in this study are shown 
through Aitoff  spherical projection diagrams in Galactic  coordinates (longitude $l$, latitude $b$) within the $l=[-90^\circ,+90^\circ]$ interval.

\item
We order bins   by density value. The main over-density region is identified 
around
the highest value bin. Subsequent  over-densities are identified by selecting the next bin,
in order of decreasing density,
 which is separated more than 15$^\circ$  from the center of all the previously defined over-densities. In this way
over-density regions  
  are differentiated and isolated. 
  For each of these regions, the midpoint of the   highest-density bin  will define the corresponding 
  \textit{ density peak}'s  coordinates.

\item
We quantify how much a certain satellite $s$ has contributed to a given density peak $p$ (which we refer to as  '$C_{p, s}$'). 
To this end, we define an aperture angle of 15$^\circ$  around the density peak, selecting all 4-galaxy-normals within it. 
For each of them, the four contributing satellites  are determined.
A given satellite $s$ is counted to contribute  the 4-galaxy-normal's \textit{weight} to peak $p$.
Therefore, its final contribution $C_{p, s}$, is 
the sum of weights corresponding to  all the  4-galaxy-normals within the peak aperture that satellite $s$ contributes to.
This has been normalized using  $C_{N, all}$, the total weighted number of 4-galaxy-normals, included those that are not within 15$^{\circ}$ of some peak center.
%%%, such that the sum $ \sum_{p, s} $C$_{\rm p, s}=1$ holds. 
Such normalization is necessary to allow for a meaningful comparison of results coming from samples with different sizes $N_{\rm tot}$.

Finally,  all satellites are  ordered by decreasing  $C_{p, s}$ to the density peak $p$, such that the first satellite is that which contributes most.

We note that changing the bin size used in our analysis does not modify our results, as we find the same overdensity regions, and final order of satellites by  $C_{p, s}$.
%While it does slightly change the position found for the density peak centres, the differences are small and do not modify the final order of satellites by  $C_{p s}$. Therefore the final results remain unaltered.

\end{enumerate}

\subsection{An extension to the method: peak strength and plane quality analysis}
\label{PMExt} 

To allow an individual and in-depth analysis of each overdensity and its corresponding predominant planar structure of satellites, we present an extension of the  4-galaxy-normal density plot method.

On one hand,
to each peak $p$ we assign
a number, $C_p$, the peak `strength', defined as the normalized number (or \%) of 4-galaxy-normals within 15$^\circ$ of the respective peak  center; that is $C_p \equiv \sum_s C_{p, s}$,   where the contribution-number  $C_{p, s}$ of the $s$ satellite to the $p$-th peak has been  defined  above. This is a useful concept  to quantify the comparisons of different density peaks 
and ultimately study the `relevance' of given planar satellite structures over others
%with each other 
(see Paper II for more details).

On another,  rather than a plane per overdensity,
% the extension \textbf{to the method}
% will provide us
 the extended method provides   a  collection  of planes, each consisting of 
 %with 
 a different number of satellites $N_{\rm sat}$.

For each over-density $p$ 
we initially fit a plane to the $N_{\rm sat}=7$  satellites with highest $C_{p, s}$ 
(i.e., the 7 satellites that contribute most  to 4-galaxy-normals within  15$^\circ$ of the density peak),
 and store the resultant plane parameters. 
This number $N_{\rm sat}=7$  is low enough to allow for an analysis of ToI parameter behaviour as $N_{\rm sat}$ increases, and at the same time high enough that we begin with populated  planes. Note that taking instead $N_{\rm sat}=7 \pm 2$  to begin with does not alter our conclusions.

Then, the next satellite in order of decreasing 
$C_{p, s}$
 is added to the group of satellites. Again a plane is fitted to their positions and the parameters stored.
This plane-fitting process is repeated
until all contributing satellites are used.

To include the effect of distance uncertainties, in practice we calculate 1000 random positions per satellite, and fit 1000 planes at each iteration with $N_{\rm sat}$ satellites.
The final results at each $N_{\rm sat}$ 
 correspond to the mean values from these random realizations  and the corresponding  errors to the standard deviations.

In this way, for each over-density found  we obtain a  \textit{collection} or catalog of
planes of satellites, each plane consisting of
% an increasing 
a different 
number of members increasing from 7 to $N_{\rm tot}$,
as well as the quality indicators for each of them.

In this work ``high quality" means  populated and
flattened planes.
This is quantified through $N_{\rm sat}$ and 
$c/a$ 
\citep[and/or $\Delta$RMS,
but note that they are very often correlated; see][]{Pawlowski14b}. 
 Being a two-parameter notion, to  compare 
planes' qualities  we need that  either $N_{\rm sat}$ is constant or that $c/a$ is constant (or that at least it varies very slowly with $N_{\rm sat}$). 
In the first case, lower $c/a$ means higher quality. In the second case, more populated planes are rated as of higher quality.  Another case when comparison 
is possible is when one plane is more flattened and populated than another: the first has a higher quality than the second. 
These considerations have been applied to the different member planes in the collection obtained for each density peak, allowing us to make quality comparisons, 
in particular with already determined planes,  and, very interestingly,  to single out new high-quality ones.\footnote{  Note that quality as understood in this paper should not be confused with
%%% `significance'
the different statistical concepts linked to the probability of occurrence of a particular plane of satellites in a cosmological simulation or  in randomized satellite systems.  }

We finally note that the advantage of the 4GND plot method lies in that it selects the subsamples of $N_{\rm sat}$ satellites with lowest $c/a$ provided that they are embedded in an underlying more populated, predominant planar structure. 
In the case of low $N_{\rm sat}$ relative to $N_{\rm tot}$ planes, this means it avoids choosing spurious structures, that although extremely narrow, are most probably not physically relevant and have a higher chance of being just a chance alignment.
For high  $N_{\rm sat}$ relative to $N_{\rm tot}$, it means the method selects the subsample of satellites that yields the highest-quality (i.e., thinnest) plane out of all possible combinations of $N_{\rm sat}$  satellites.

\section{Results for the MW}
\label{ResultsMW}

Two different satellite samples, MW27 and MW46, have been analyzed for the MW, see Section~\ref{observedplanes} for their specification. 
The first of them had been analyzed previously 
(see P13)
%\citep[see][]{Pawlowski13} 
and we use it as a reference to underline how the extended method finds new, high quality planes. MW46 compiles the to-date confirmed satellites of the MW 
%%%as of October 2019, 
(see Table~\ref{tab:mwdata}). Comparing the results found for each  of them is interesting because it allows to quantify  the extent to which the added satellites follow the planar structures already found with the MW27 sample,
and assess the robustness of such spatial structures.

\begin{figure}
\centering
\includegraphics[width=\linewidth]{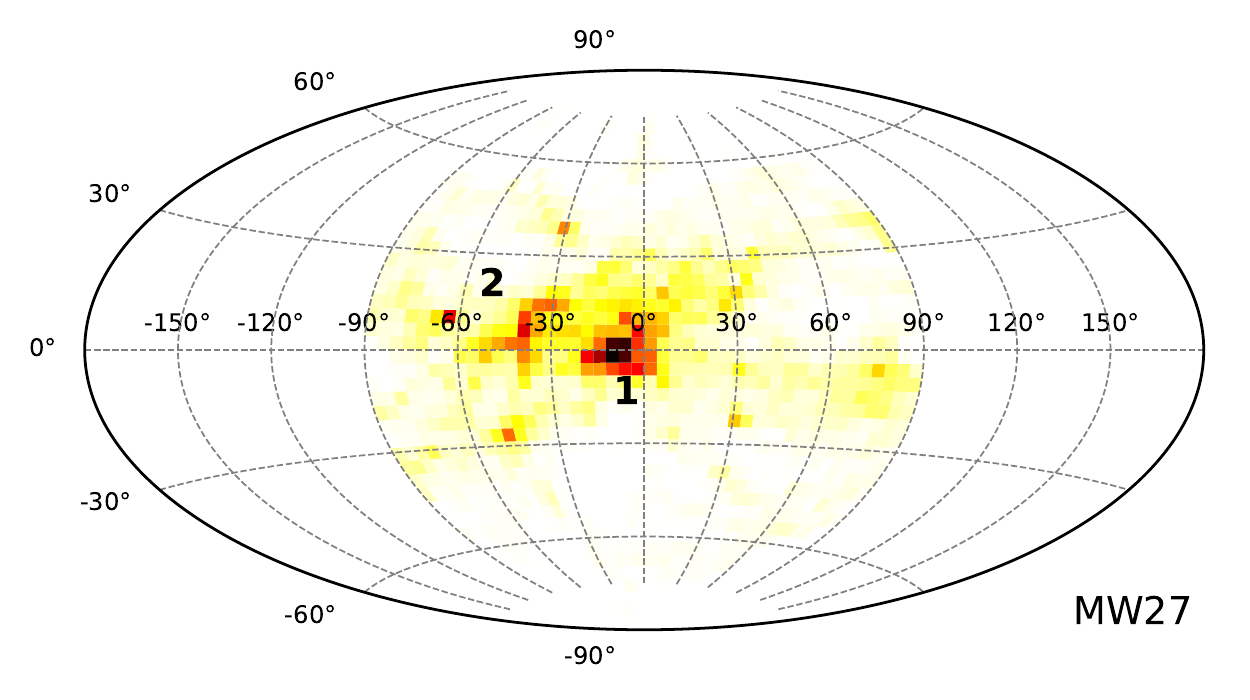}\\
\includegraphics[width=\linewidth]{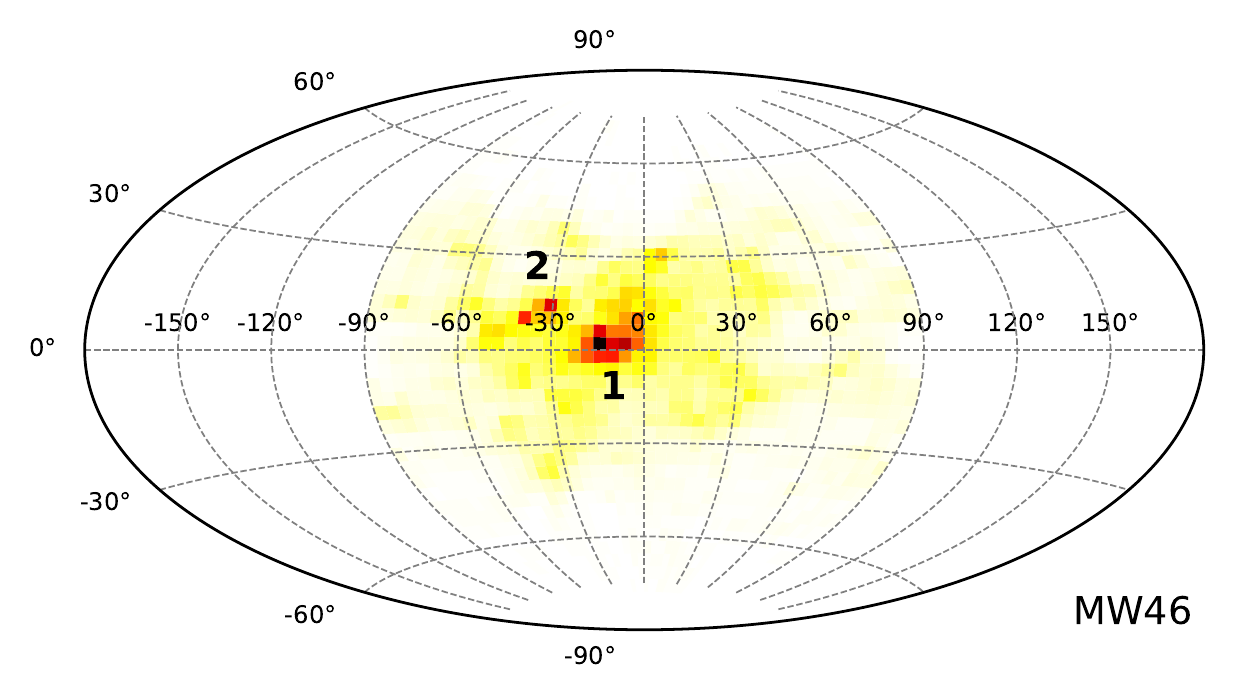}\\
\includegraphics[width=\linewidth]{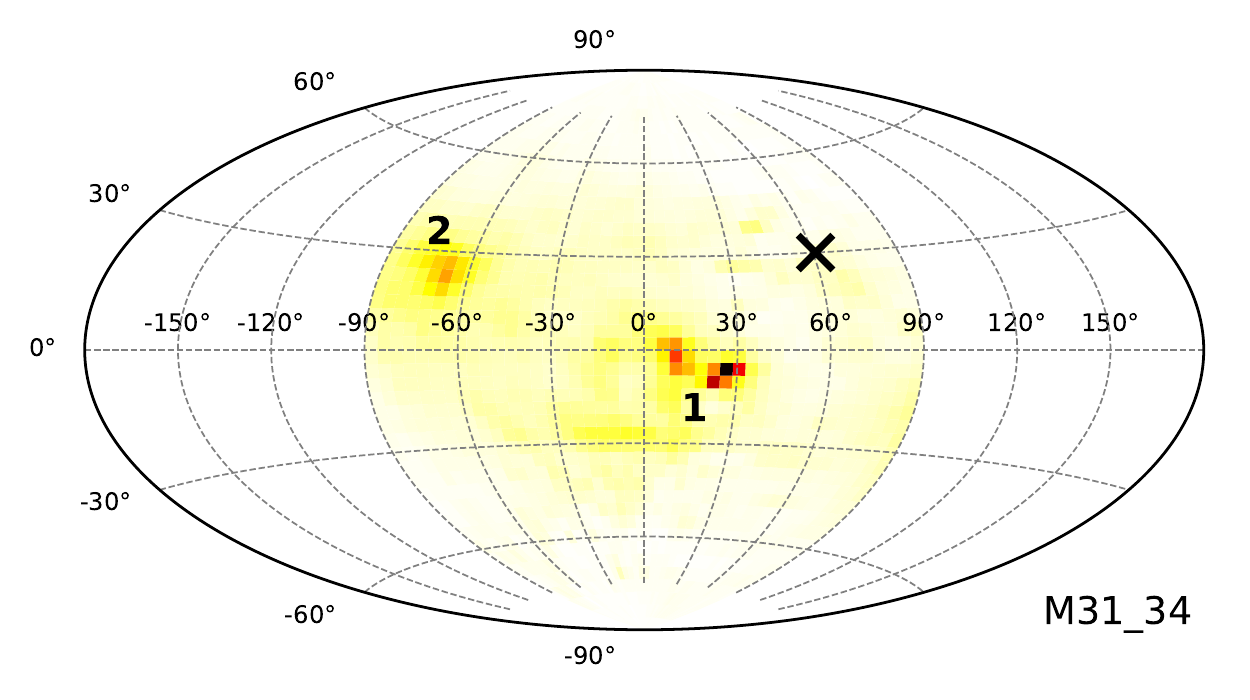}\\
\caption{Aitoff projection diagrams of the Milky Way   MW27 (top), MW46 (middle)  and M31\_34 (bottom) 4-galaxy-normal density plots
\citep[see also figs. 2 and 4 in][]{Pawlowski13}.
The colormap shows the 
number of 4-galaxy-normals within a bin, each weighted by $\log\left( \frac{a+b}{c} \right)$
to emphasize planar-like spatial configurations of satellites
(see \Sec{N4p} for details).  
The total number of 4-galaxy-normals  is 1,755,000   for MW27,
%\textbf{6604500 for MW37}
16,318,500 for MW46
  and  4,637,600 for M31\_34, taking into account 100 random realizations of satellite radial distances within the errors for each (see text).
The relevant over-density regions in each map are labeled in order of intensity (Peaks 1 and 2 in the main text).
%Each diagram is centered on its corresponding central galaxy but the   orientation of coordinates in both cases is such that the disc of the MW lies on the latitude $b=0^\circ$ plane and its spin points along the OZ axis.
Diagrams are in Galactocentric coordinates, centered in the Galactic center.
See Tables~\ref{tab:mwdata} and \ref{tab:m31data} for details on the different MW and M31 satellite samples and their data.
M31's spin is marked with an `X'.
%A much finer bin size than that shown here has been used to extract the density peak coordinates.
All three diagrams share a common colorbar with values   proportional to the normalized  bin density.
}
\label{MWM31dp}
\end{figure}

%%%%%%::::::::::  MW27  BAR CHART  :::::::::::::::::

\begin{figure*}
\centering
\includegraphics[width=\linewidth]{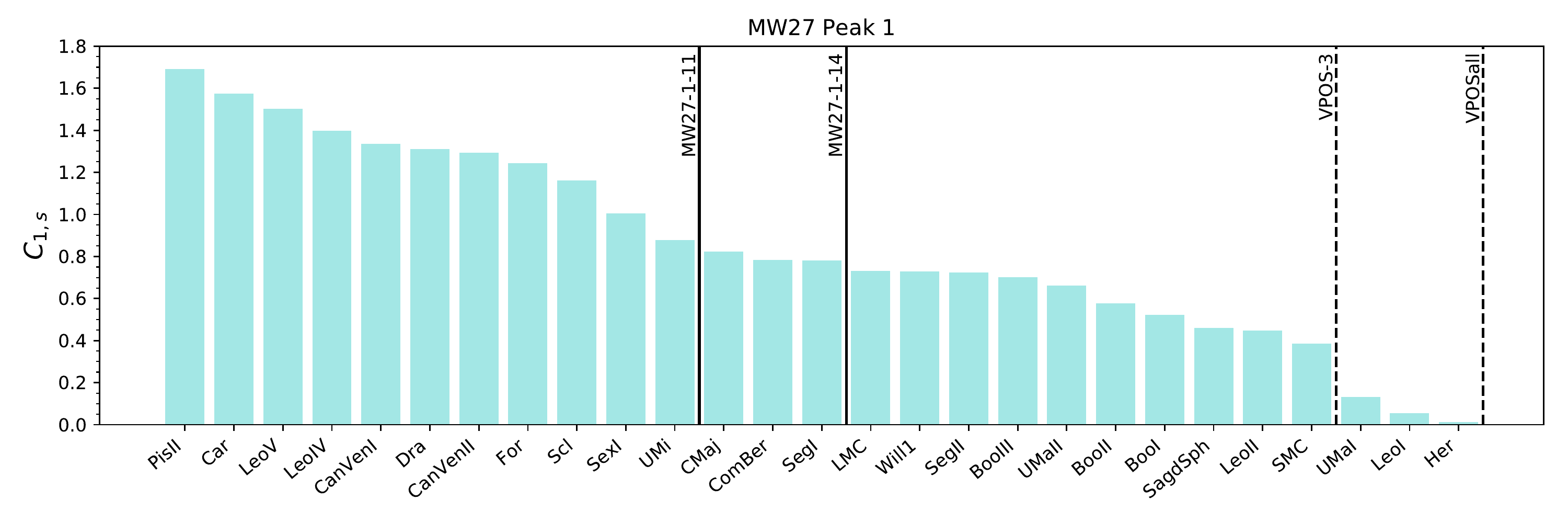}
\includegraphics[width=\linewidth]{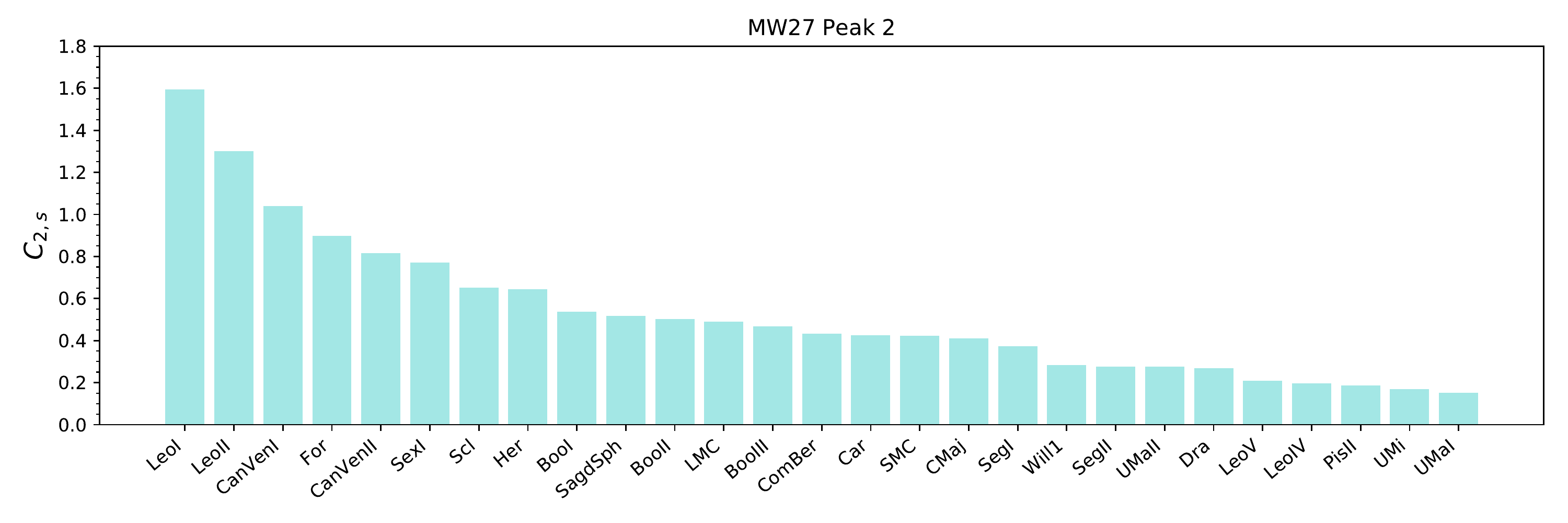}\\
\vspace{-0.4cm}
\caption{Bar chart showing the weighted-contribution 
of  satellites to 4-galaxy-normals in 15$^\circ$  around the first  ($C_{1,s}$, top panel) and second ($C_{2,s}$, bottom panel) most important  over-densities found in the Milky Way MW27 density plot (upper panel in \Fig{MWM31dp}).
$C_{p,s}$ values are normalized using  $C_{N, all}$, the total weighted number of 4-galaxy-normals, included those that are not within 15$^{\circ}$ of some peak center.
The sets of objects that make up the planes of satellites singled out in this work are delimited with vertical lines and labeled correspondingly (see Table~\ref{table_obsThisWork}).
Planes quoted in the literature, in particular \citet{Pawlowski13}, are marked with dashed lines.
}
\label{barMW}
\end{figure*}

%%%%%%::::::::::  MW NEW SAMPLE BAR CHART  :::::::::::::::::

\begin{figure*}
\centering
\includegraphics[width=\linewidth]{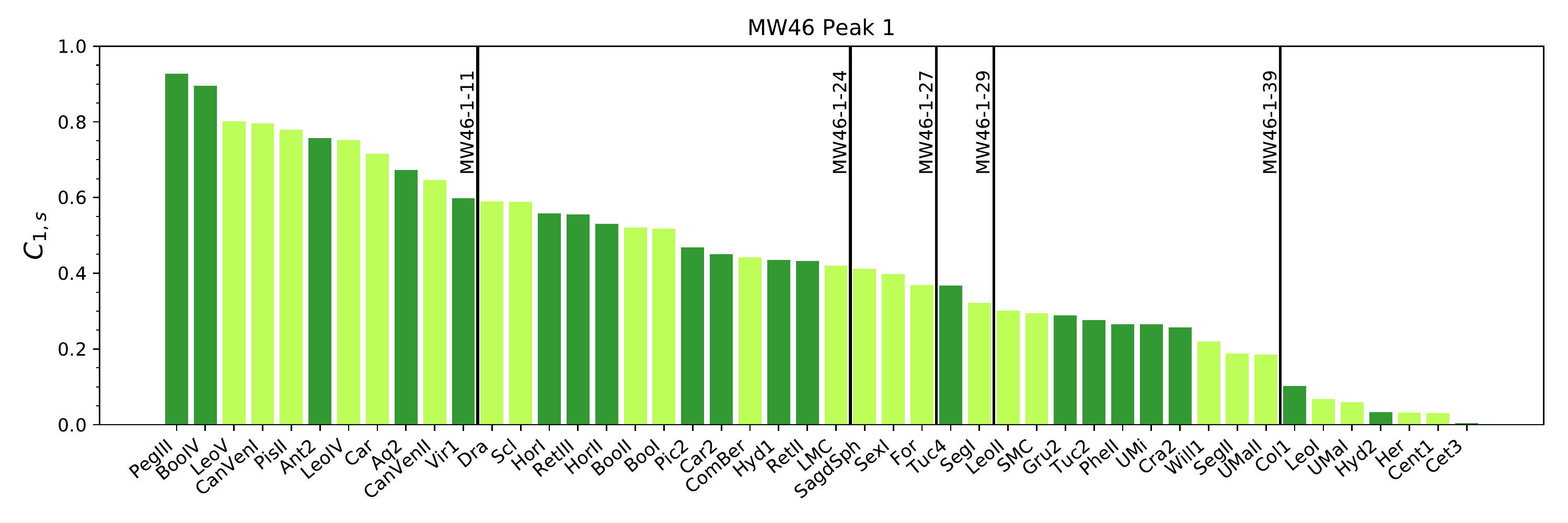}
\includegraphics[width=\linewidth]{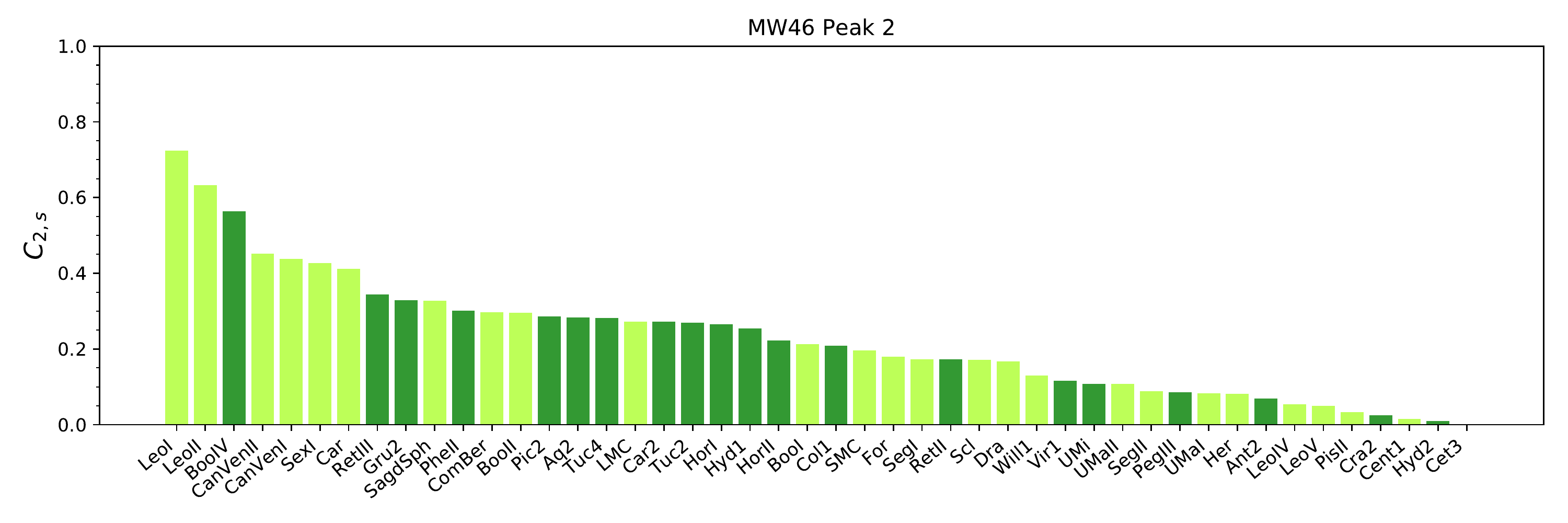}\\
\vspace{-0.4cm}
\caption{Same as Fig.~\ref{barMW} for the MW46 sample. 
The contributions from new satellites included in  MW46 with respect to MW27 are plotted in a darker color.
Note that many of the new satellites considered just fill up the structures already discovered with the MW27 sample (see samples in Table~\ref{tab:mwdata}). 
%%\magenta{Most satellites belonging to both MW27 and MW46 samples do not appreciably change their contribution to M31 peaks.  Bootes II, Bootes I, SagdSph, Leo II and the SMC have increased their contribution to Peak 1 relative to their contribution in the MW27 sample}
}
\label{barMW46}
\end{figure*}

%%%%%%::::::::::  M31  BAR CHART  :::::::::::::::::

\begin{figure*}
\centering
\includegraphics[width=\linewidth]{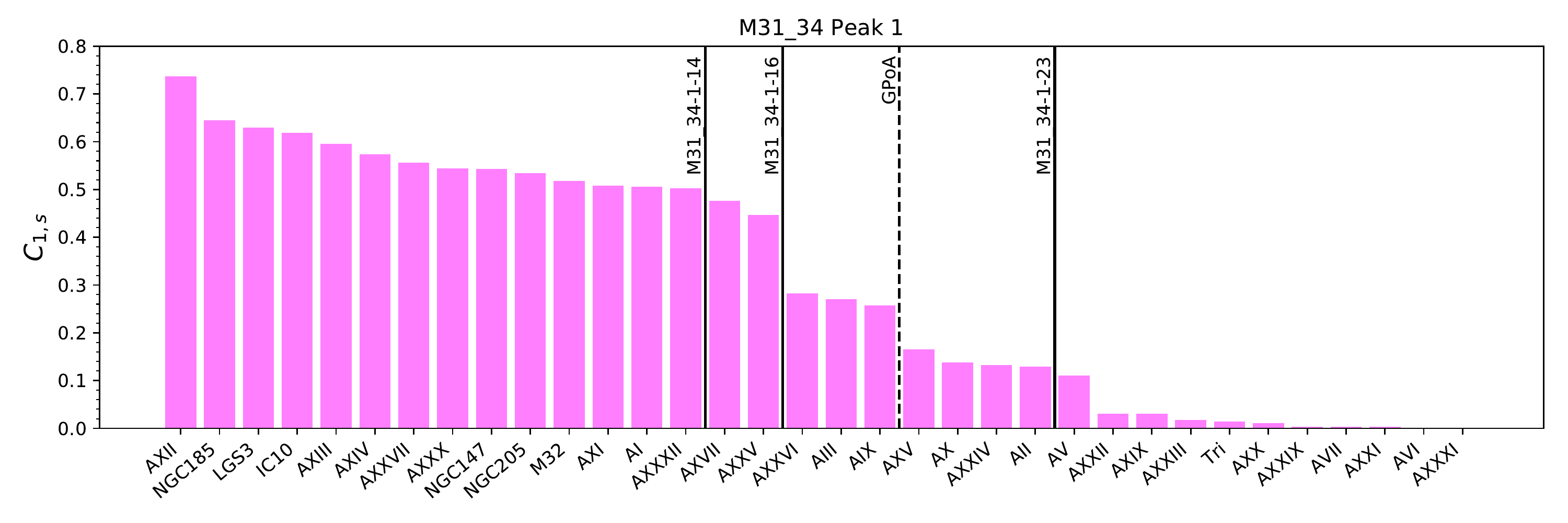}
\includegraphics[width=\linewidth]{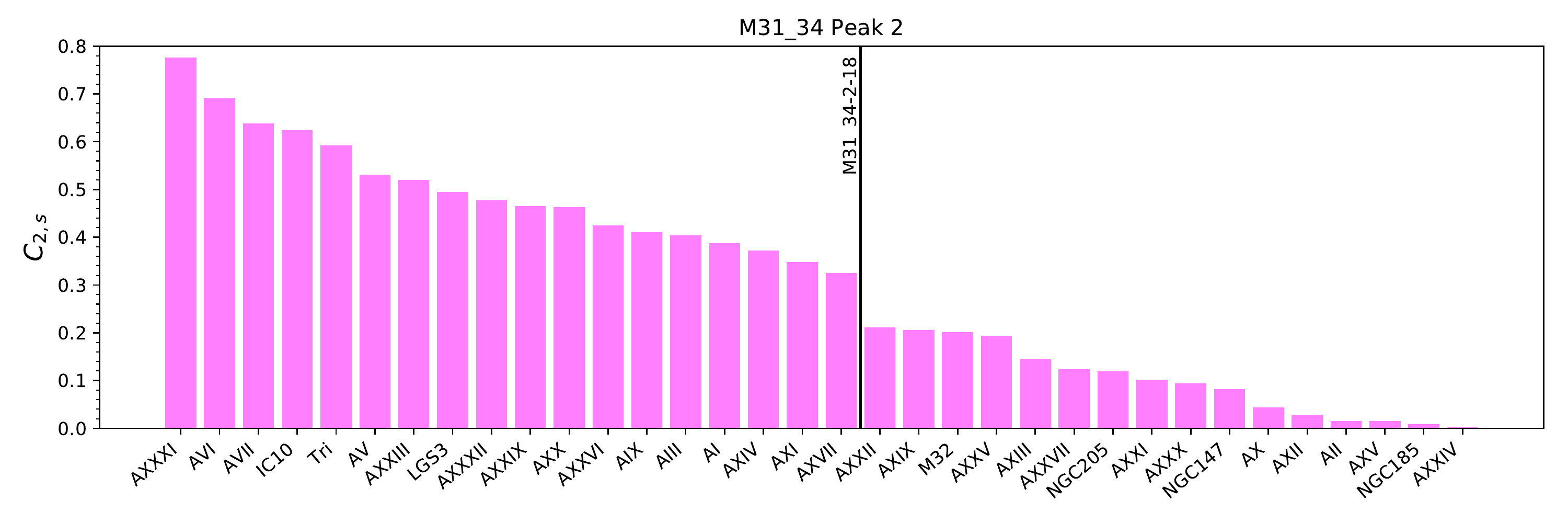}\\
\vspace{-0.4cm}
\caption{Same as \Fig{barMW} for M31\_34. Both the first (top panel) and second (bottom panel) over-densities of M31's 4-galaxy-normal density plot have similar strengths (see text and Table~\ref{c1mwm31}) which is reflected in a high and comparable contribution from satellites in both bar charts.
}
\label{barM31}
\end{figure*}

\subsection{The 4GND plot for the MW27 satellite sample}
\label{MW}
The upper panel of   \Fig{MWM31dp}  shows the Aitoff projection diagram of the  4-galaxy-normal density plot obtained for the MW27 satellite system.
 This Aitoff diagram can be compared to  the contour plot in  fig.~2 from  
 %\citet{Pawlowski13},
 P13, to which it is  essentially identical\footnote{
The conversion from the  galactocentric longitude convention used here ($l$, centered in the Galactic center) and that used  in 
\citet{Pawlowski13} ($l'$, centered in the Galactic anticenter) is: $l = l' -180 (^\circ)$.
}.

As reported in 
%\citet{Pawlowski13},
P13,
the density plot for the MW27 sample shows that 4-galaxy-normals are mainly clustered in the region central to the diagram,
revealing a planar structure that is polar to the Galaxy (i.e., the normal vector to the plane is perpendicular to the Galaxy's spin vector).
There is one dominant over-density (Peak 1)
located at 
%\red{ $(l,b)=(-10.23, -0.41)$,}
$(l,b)=(-10.2^\circ, -0.4^\circ)$,
 and  a second lower density peak (Peak 2) close to the first
 at
%\red{ $(l,b)=(-39.68, 4.50)$. }
 $(l,b)=(-39.7^\circ, 4.5^\circ)$. 
 %%We will neglect the few isolated bins with intermediate intensity.

Fig.~\ref{barMW} shows MW27 satellites ordered by $C_{p,s}$ contribution
to 4-galaxy-normals within 15$^\circ$ around both density peaks  (i.e., $C_{1,s}$, upper panel, and  $C_{2,s}$, bottom panel).  
There are 11 satellites  that dominate the contribution
 (i.e., $C_{p, s} > 0.5 \times {\rm max}(C_{p, s}, s = 1, ..., 27$)), 
 in Peak 1 (i.e., PiscesII, Carina, LeoV,  LeoIV, CanesVenaticiI, Draco, CanesVenaticiII,  Fornax,  Sculptor, SextansI, UrsaMinor),
  while 4    contribute most in Peak 2 (i.e., LeoI, LeoII, CanesVenaticiI, Fornax)\footnote{Both the location of these peaks  and their major contributers are consistent with the results reported in  \citet{Pawlowski13} (see their fig.~3).}.
  In this case, the contribution is mainly driven by LeoI and LeoII, while the rest of satellites are common with Peak 1 and take low $C_{p, s}$ values, indicating the low relevance of this second structure.

\subsection{The 4GND plot for the MW46 satellite sample}
\label{MW-46}

In the  middle  panel of Fig.~\ref{MWM31dp} we plot the Aitoff projection diagram for the  4-galaxy-normal density plot of the MW46 satellite system. 
This density plot is very similar to that for the   MW27 sample (above).
%%% Comparing with that of the MW27 \textbf{(above),} we see that both are very similar. 
Indeed, Peaks 1 and 2  are placed at 
%\red{$(l,b)=(-14.31, 2.04)$ and  $(l,b)=(-30.68, 14.32)$} ,
$(l,b)=(-14.3^\circ, 2.0^\circ)$ and  $(l,b)=(-30.7^\circ, 14.3^\circ)$,
 respectively, close to the MW27 main peaks location.
In both cases there is a dominant peak and a secondary one, with no other relevant structures.
To make the comparison quantitative, in Table~\ref{c1mwm31} we give the values for $C_{\rm MW27, p}$ and $C_{\rm MW46, p}$, with $p$=1,2 for each peak. 
%%We can see that the  MW27 and MW46 peaks strengths are similar.
The peak strengths are somewhat lower for both peaks in the case of MW46 than MW27, as expected given the higher number of satellites and therefore of 4-galaxy-normals.

%%%%%%%%%%%   TABLA VALORES  C1 %%%%%%%%%%

\begin{table}
\centering
\begin{tabular}{l l l }
\hline
 & $C_1 \pm \sigma$  & $C_2 \pm \sigma$ \\
 \hline
 \hline
 MW27 & 22.92$\pm$0.26 &  14.31$\pm$0.20 \\
 MW46 & 19.59$\pm$0.19 &  10.49$\pm$0.13 \\
 %%%M31  & 10.53$\pm$0.62 &  10.53$\pm$1.62 \\
 M31\_34  & 10.53$\pm$0.62 &  10.52$\pm$1.62 \\
 M31\_36  & 11.89$\pm$0.59 &   9.80$\pm$1.85 \\
 \hline
\end{tabular}
\caption{Peak strengths $C_p$  for the main 2 peaks found in the MW27, MW46, M31\_34  and M31\_36 4GND plots.
Peak strength is computed as
$C_p \equiv \Sigma_s C_{p,s}$, and is a percentage, as $C_{p,s}$ is normalized to the total weighted number of 4-galaxy-normals.
 Errors are calculated from 100 random realizations using the radial distance uncertainties.}
\label{c1mwm31}
\end{table}

%%%%%%%%%%%%%%%%%%%%%%%%%%%%%%%%%%%%%%%%%%%

Bar charts showing the satellite contribution $C_{p,s}$ to density peaks 
for the MW46 sample are given in  Fig.~\ref{barMW46}.
The contributions from the 21 new satellites that are not in MW27 have been marked with a darker color.
%, with  $C_{1s}$, upper panel, and  $C_{2s}$, bottom panel. 
Comparing to Fig.~\ref{barMW}, we see that 
i), most satellites belonging to both MW27 and MW46 samples do not appreciably change their contribution to MW peaks, and ii),  
several satellites among the  21  that are not in MW27 have a relevant contribution to Peak 1 of  MW46.
An interesting result is that the 25\% of satellites that contribute the most to MW27 Peak 1 are included among the 25\% of satellites that contribute the most to MW46 Peak 1, complemented with 5 satellites that are not in MW27.
Also, out of the 24  satellites that dominate the contribution to Peak 1, 12  are unique to the MW46 increased sample.   
%And out of the 11 satellites that contribute most, 5 are unique to MW46.

As for Peak 2, among the 6 satellites in MW27  contributing the most, all but Fornax are among the 25\% main contributors to MW46 Peak 2. And out of the 24  (11) satellites that dominate the contribution to Peak 2, 12  (4) are unique to the MW46 sample.

In summary, the new MW confirmed satellites seem to just add   to the  positional planar structures already determined with 27 satellites. 
In the next section this point will be made quantitative through the comparison of the directions of the normal vectors to the different planes. 
%%\green{NO HE DICHO NADA     DE LOS 5 SATELITES QUE INCREMENTAN SU CONTRIBUCION AL PICO 1}

\subsection{Quality analysis}
\label{Quality_MW}

Following our extension to the 4-galaxy-normal method  (\Sec{PMExt}),  for each over-density region
  we have iteratively computed planes of satellites with an increasing number of members $N_{\rm sat}$,
  following the order of satellites given in Fig.~\ref{barMW}.
In this way,
 for each peak we obtain an ordered collection of planes, one plane for each $N_{\rm sat}$. 
These collections, and the identities of the  $N_{\rm sat}$ satellites belonging to each plane, can be read out of Figs.~\ref{barMW} and \ref{barMW46}, for both peaks of the MW27 and MW46 samples, respectively.

\begin{figure*}
\centering
\includegraphics[width=0.49\linewidth]{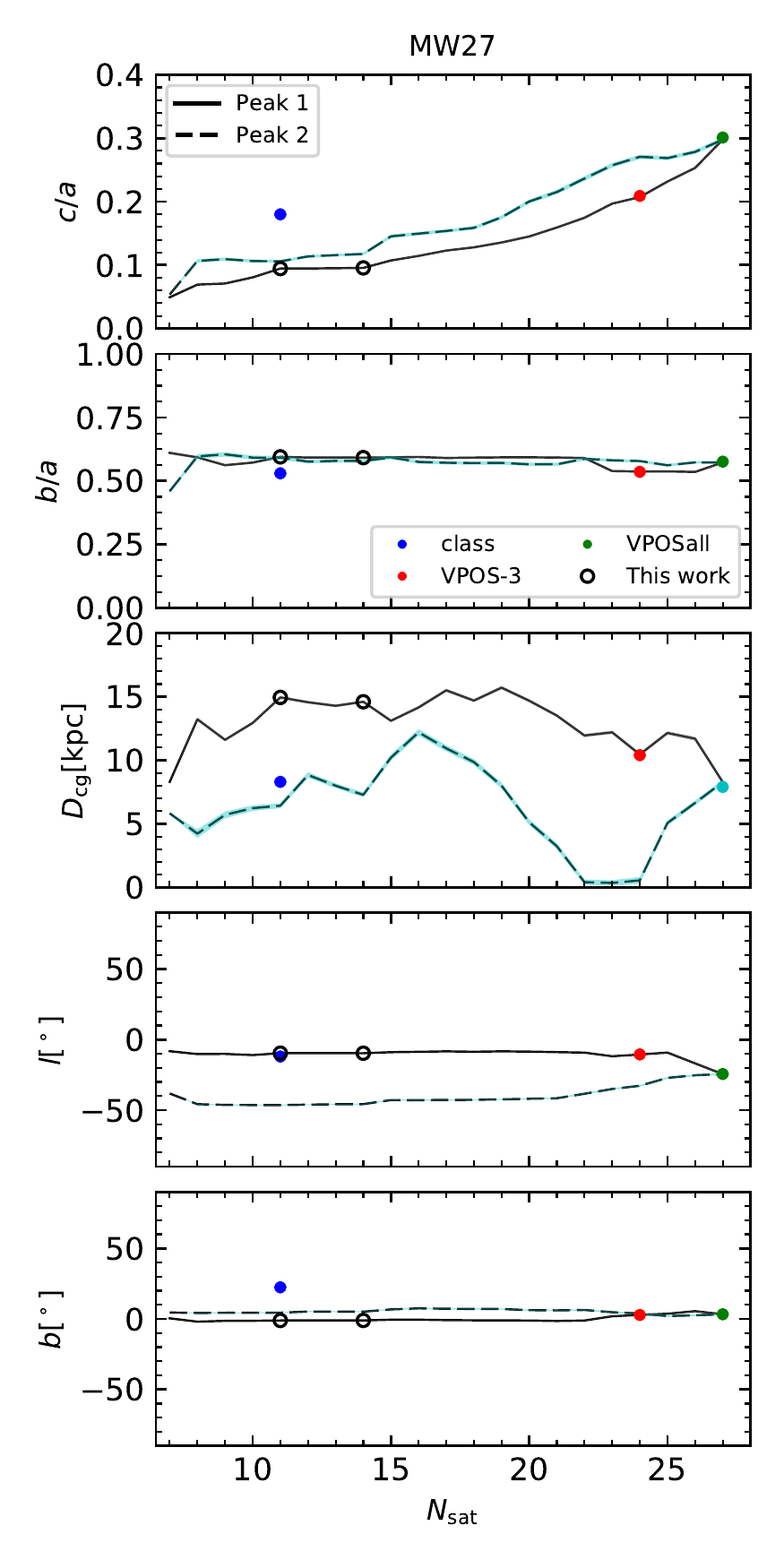}
\includegraphics[width=0.49\linewidth]{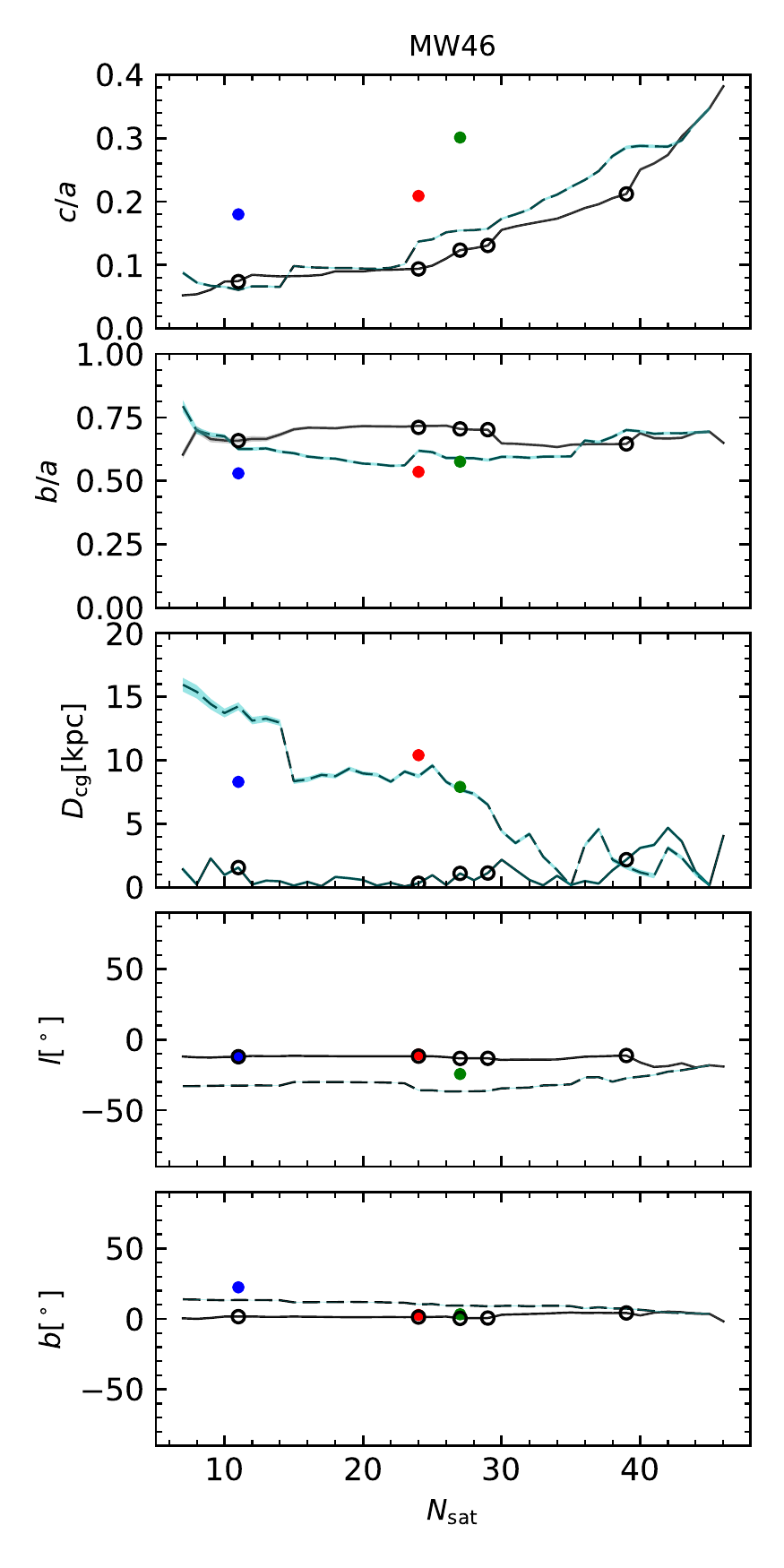}
\vspace{-0.4cm}
\caption{Quality analysis of the main planar structures found in the Milky Way with the 4-galaxy-normal density plot method (see Fig.~\ref{MWM31dp}). \textit{Left:} \citet{Pawlowski13}'s sample with 27 satellites. \textit{Right:} the updated MW46  sample with 46  satellites.
Lines show   $c/a$, $b/a$,
%% $\Delta$RMS and 
$D_{\rm cg}$ and
the direction of normal vectors to the best-fitting plane $(l,b)$ as a function of $N_{\rm sat}$.
A solid line illustrates results for the Peak 1 planar structure, and a dashed line those for Peak 2.
 Except in the  $l$ or $b$ versus $N_{\rm sat}$  panels, 
results are the mean values of 1000 realizations at each $N_{\rm sat}$, and shaded regions
(gray for Peak 1 and cyan for peak 2)
 show the standard deviations.
Note that these are very small.
 $(l,b)$   directions have been calculated using the most-likely positions of satellites  and shaded regions show the corresponding uncertainties
in terms of the spherical standard distance $\Delta_{\rm sph}$ \citep{Metz07}.
As expected, the two curves in each panel converge for the maximum $N_{\rm sat}$ since the samples of satellites
considered for each Peak become identical.
Colored circles
show the results for the reported observed planes of satellites in the MW (classical, VPOS-3, VPOSall).
%%%  and in M31 (`Ibata-Conn-13' and GPoA), respectively.
 Their specific values including their errors are given in \Tab{table_obspl}.
The planes of satellites singled out in this work are shown with black open circles, and their corresponding  ToI parameters 
are given in \Tab{table_obsThisWork}.
}
\label{MWM31_ca_frac}
\end{figure*}

The left panel of 
\Fig{MWM31_ca_frac} shows the 
%%%%\textbf{plane characteristics} results obtained for 
characteristics of the planar
 structures defined by  
Peak 1 (solid line) and Peak 2 (dashed line) in the MW27 sample.
We focus on $c/a$, $b/a$, $D_{\rm cg}$ and the normal direction to the plane. 
Quantitative  results of RMS heights are provided 
in tables
but not shown in our figures as this parameter correlates with $c/a$ presenting the same general trends as   $N_{\rm sat}$ increases, and not adding any other relevant information.

The collection of planes obtained for a given over-density region gives rise to a set of points in this plot, 
with their corresponding error bars, one at each $N_{\rm sat}$ value.
 They are shown joined with a line, with the corresponding error bands.
  For Peak 1, errors are shown as a grey shade; for Peak 2 as a cyan shade.
 These are very narrow 
 %in this case
(even imperceptible in some cases), showing that the MW27 results are hardly affected by distance uncertainties.

First we see that, for any $N_{\rm sat}$ value, $b/a$ is rather high (and constant), while $c/a$ is low. Therefore configurations are indeed  \textit{planar-like}. Moreover,
the  lines for both density peaks show rather smooth trends of increasing $c/a$ 
%and $\Delta$RMS 
with increasing $N_{\rm sat}$.
In particular,
 the MW27 Peak 1 line 
 defines a planar structure  of  satellites with a higher quality (i.e., lower $c/a$
 %,  lower $\Delta$RMS
 % (not shown in the Figure)  
 at given $N_{\rm sat}$), than that defined by Peak 2. 
This is expected, given the 
%higher density of 4-galaxy-normals in 
higher  $C_p$ strength of
Peak 1 than  Peak 2
(see Fig.~\ref{MWM31dp} and Table~\ref{c1mwm31}).
This suggests
that the MW27 satellite sample seems to be a unique highly planar-like  organized  structure, as we had already learnt from Fig.~\ref{barMW}.
%We also note that both the $c/a$  and $\Delta$RMS   versus $N_{\rm sat}$  curves  roughly show  the same shape patterns, due to the lack of significant $b/a$ variation as   $N_{\rm sat}$ increases. Therefore, including $\Delta$RMS on top of $c/a$ in the quality analysis generally  does not add any relevant information.

In the bottom panels we give the directions of the normal vectors corresponding to the best fitting planes (obtained from the satellites' most-likely positions) as a function of $N_{\rm sat}$. Shaded regions show the
 corresponding spherical standard distances $\Delta_{\rm sph}$ \citep{Metz07},
a measure of the
 collimation of plane normals
 %planes normals' collimation 
  for the 1000 realizations.
The normal directions to planes from both Peak 1 and 2 in  MW27 remain very stable as  $N_{\rm sat}$ increases, and their uncertainties due to satellite distance errors are very small.

As for the plane distances to the MW, $D_{\rm cg}$ is  below 16 kpc in all plane-fitting iterations in Peak 1, and below 12.5 kpc in the case of Peak 2.

For reference, overplotted
colored points in all panels show the values on this diagram for
the  observed planes of MW satellites mentioned in the literature (classical, VPOS-3, VPOSall; see \Tab{table_obspl}),
defined from Peak 1
(P13). 
%\citep{Pawlowski13}.
We note that the points corresponding to the VPOS-3 and VPOSall planes  fall over the trend given by the solid  line, as expected.
  On the other hand, and very interestingly, 
 this analysis    shows that 
    there  is  a  different 
 combination of $N_{\rm sat}$=11 satellites that results in a much flatter and thinner plane than the classical one
(see MW27-1-11\footnote{Planes underlined in this work, either for the MW or M31, are named after the peak where they have been identified and the number of satellites they include (i.e., [Sample-Peak-$N_{\rm sat}$]). } in Table~\ref{table_obsThisWork}).
 In fact, the plane including $N_{\rm sat}$=14 satellites (MW27-1-14 in Table~\ref{table_obsThisWork})
presents an even higher quality than that with $N_{\rm sat}$=11, as  
$c/a$  remains  roughly constant at a higher $N_{\rm sat}$.

This is possible
because this analysis uses  the 3-dimensional information of positions, while the classical plane of satellites
 was found observationally
  when only the most luminous (i.e. massive) satellites were known to exist.
   This important result
 indicates
 that planes of satellites are not necessarily  composed by the most  massive satellites of a galactic system
  \citep[see also][]{Libeskind05,Collins15} 
  and hence they should not be searched for in this way.
  Indeed, 
as we show in Fig.~\ref{cps_mstar}, 
  we find that $M_{\rm star}$
is not correlated with $C_{1,s}$ (contribution to the main over-density region, where we find the highest quality planes):
% to a higher than 76\% confidence level.
%(log10, log10)     -0.06174275645649642     0.7596535819427815
 the correlation coefficient $r$  is low, in such a way that the probability of getting such value assuming that there is no correlation is higher than 76\% (see upper panel in Fig.~\ref{cps_mstar} for MW27 results).

In the right panels of Fig.~\ref{MWM31_ca_frac} we plot the results for the MW46 sample. It is very interesting to see that they do not differ that much from those of the MW27 sample, despite the MW46 sample contains 21 satellites that are not members of MW27.

As before,  we find 
planes of satellites with higher qualities than those previously reported  with a given $N_{\rm sat}$. For example that with $N_{\rm sat}$ =11, similar to MW27-1-11.
More specifically, 
%examples have been found 
 planes  with $N_{\rm sat}$ =24 and 27 present 
%with 
lower $c/a$ values than the VPOS-3 or VPOSall, respectively. These are marked as MW46-1-24 and MW46-1-27 in Figs.~\ref{barMW46} and \ref{MWM31_ca_frac}
(see Table~\ref{table_obsThisWork} for the values of their ToI parameters).

In Fig.~\ref{MWM31_ca_frac}  we can see that the slope of the $c/a$ versus $N_{\rm sat}$ curve grows
relatively faster 
 for more than  39 satellites. We  hence  single out the MW46-1-39 plane  with $c/a = 0.212\pm0.002$  as the most populated plane that thin identified so far in the Local Group. 
 A softer discontinuity is also apparent at $N_{\rm sat}=29$,   therefore we also emphasize the  MW46-1-29  plane as a high quality one 
 %%In Fig.~\ref{barMW37} the respective identities of their satellite members can be read, 
(see Table~\ref{table_obsThisWork}).
% for information on some of their respective  properties.
We as well find planes as thin as the VPOS-3 or VPOSall, but more populated (with $N_{\rm tot}$ = 39 and 43 satellites, respectively) and with approximately the same normal direction.

In the two bottom-most panels of Fig.~\ref{MWM31_ca_frac} we plot the $(l,b)$ coordinates of the normal vectors to the planes. 
An interesting result is that the plane with $N_{\rm sat}=24$ has the same direction as the MW27 VPOS-3 (a red circle in the plot), a direction that keeps approximately until $N_{\rm sat}=29$ (i.e., the MW46-1-29 plane), even if the identity of the satellites are not the same, with 13 satellites in MW46-1-29 not in MW27. These panels also indicate that the normals keep their directions within $3^\circ$ in $l$ and $4.5^\circ$ in $b$,
 up to $N_{\rm sat}=39$ (i.e. the MW46-1-39 plane).
These results quantitatively confirm that many of the new-discovered satellites fill in planes already delineated by members of the MW27 sample, as anticipated in the previous section.

%\textbf{Finally, the offsets $D_{\rm cg}$ from MW46 planes are smaller than those of MW27. However we note that the global differences between them is consistent with their $\Delta$RMS values. }

\begin{figure}
\centering
\includegraphics[width=\linewidth]{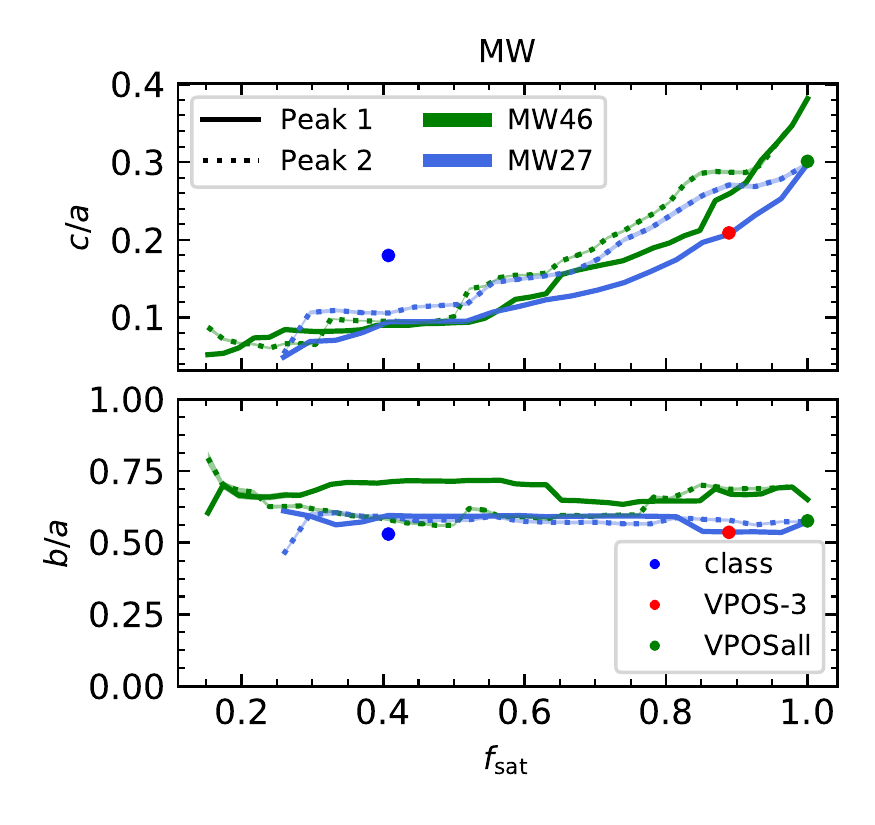}\\
\vspace{-0.6cm}
\caption{Same as Fig.~\ref{MWM31_ca_frac} in terms of the satellite fraction $f_{\rm sat}$. Blue (green) lines are results for the MW27 (MW46) satellite sample.
}
\label{MW27-MW46_fsat}
\end{figure}

The resemblances between MW27 and MW46 results can be better appreciated by plotting their properties together  in terms of $f_{\rm sat} \equiv N_{\rm sat}/ N_{\rm tot}$, see Fig.~\ref{MW27-MW46_fsat}. In both cases the quality of Peak 1 is better than that of Peak 2. Moreover, planes identified in  the MW27 sample  have, at any fixed $f_{\rm sat}$, a slightly better quality  than planes in the MW46  collection at the same $f_{\rm sat}$.  Therefore,  the main planar structure found with the sample of  the 46 currently  confirmed  MW satellites is the same as that previously found with  27 satellites, although they differ in 21  satellite members.
These results, with the current data available,  corroborate the presence of a vast polar planar structure of satellites around  the MW.

We finally note that the true census of MW satellite galaxies is currently far from complete, and thus the robustness of the results presented here requires  confirmation in the future. Indeed,
some current surveys might  bias to find new satellites near the known planar structures.
In addition,
 the advent of new deep surveys reaching increasingly fainter magnitudes and wider sky coverage (like the Rubin Observatory Legacy Survey of Space and Time, LSST), will most probably yield the discovery of a myriad of new faint dwarfs in the virial volume of the MW. The characterization of the spatial distribution of MW satellites will hence need to be progressively updated in order to corroborate  our results.
%However, the robustness of these results need to wait for confirmation until  more faint / ultra-faint satellites are discovered, BLA BLA proyectos.

%%%%%::::::::  MW27 and MW37 SAMPLES :::::::::::

%%%%%:::::::::::::: MW27 and MW37 TOGETHER     ::::::::::::::::::

%%%%%::::::::  M31  SAMPLE :::::::::::

\begin{figure}
\centering
\includegraphics[width=\linewidth]{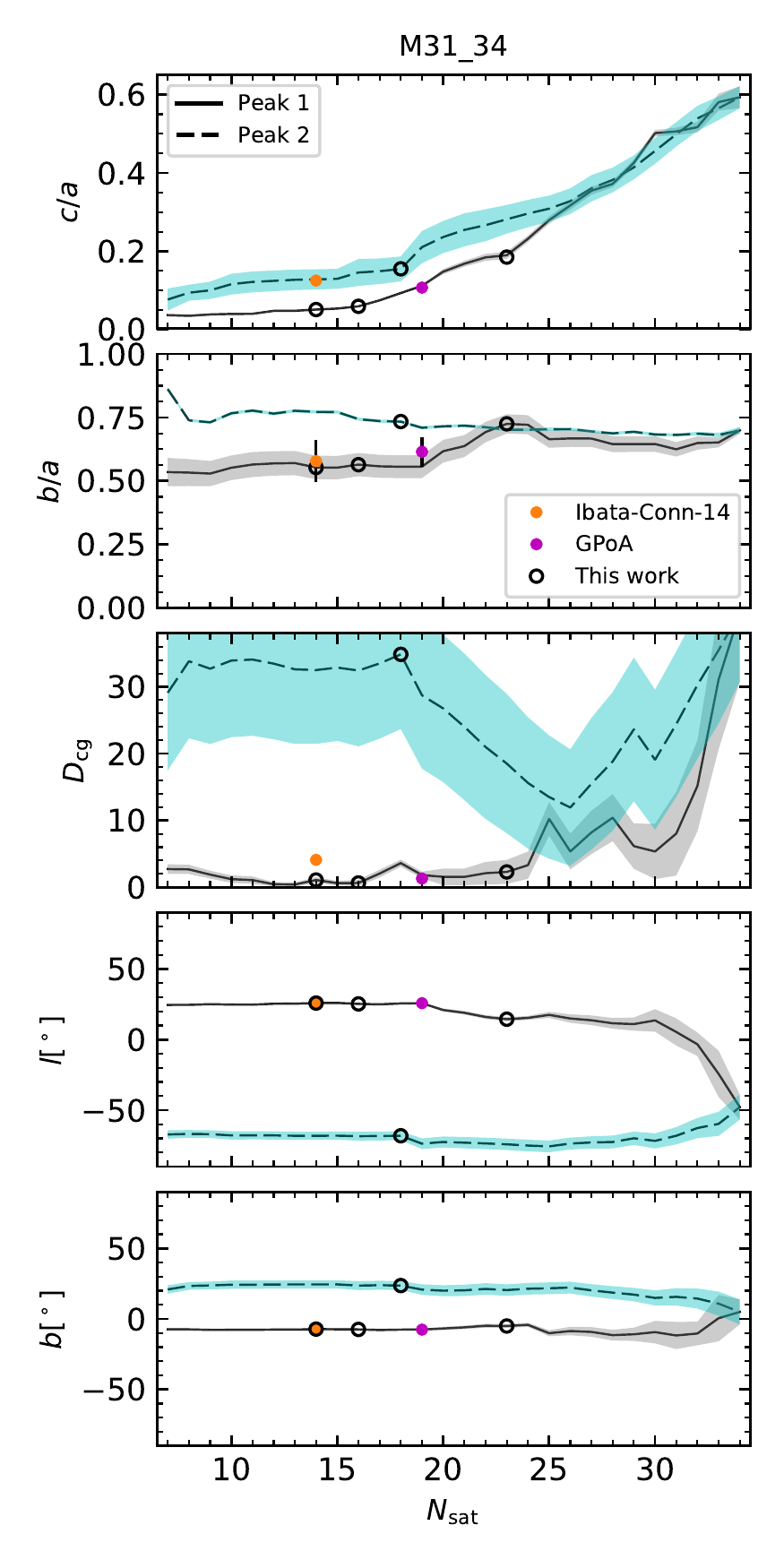}\\
\vspace{-0.6cm}
\caption{Same as Fig.~\ref{MWM31_ca_frac} for the M31\_34 satellite sample (reference satellite sample, see Table~\ref{tab:m31data}). Colored circles
show the results for the reported observed planes of satellites in M31 (`Ibata-Conn-13' and GPoA).
Their specific values including their errors are given in \Tab{table_obspl}.
The planes of satellites singled out in this work are shown with black open circles, and their corresponding  ToI parameters are given in \Tab{table_obsThisWork}.
}
\label{M31_ca_frac}
\end{figure}

%%%%%:::::::::::::::::::: Cps vs MSTAR    :::::::::::::::

\begin{figure}
\centering
\includegraphics[width=\linewidth]{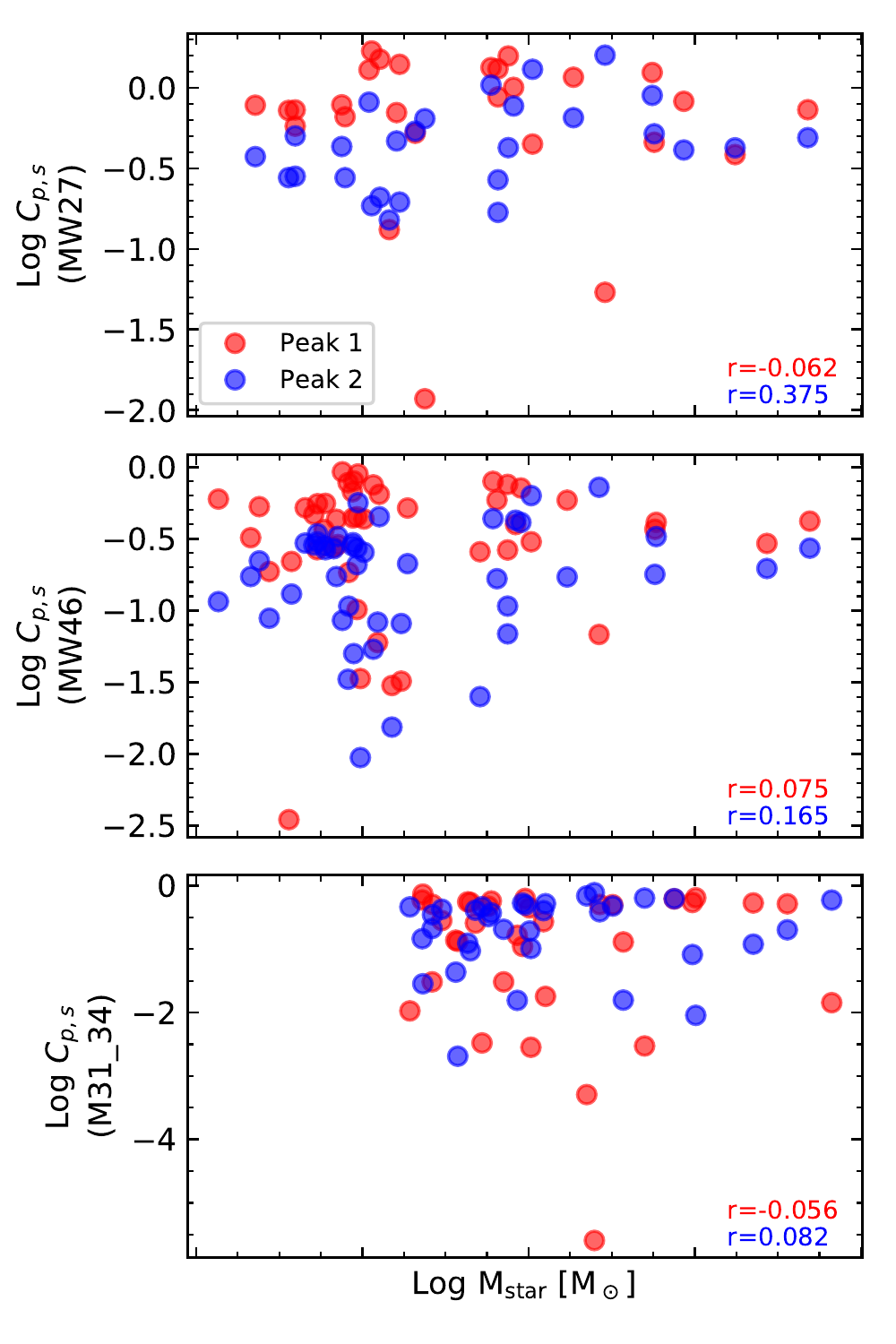}\\
\vspace{-0.5cm}
\caption{
The contribution of satellites to 4-galaxy-normals within 15$^\circ$ of the main density peaks, $C_{p,s}$, versus stellar mass. The Pearson correlation coefficients $r$ 
%%%and corresponding $p$-values 
are given in each case. 
}
\label{cps_mstar}
\end{figure}

%%%% SSSSSSSSSSSSSSSSSSSSSSSSSSSSSSSSSSSSSSSSSSS

\section{Results for Andromeda M31}\label{M31}

\subsection{The 4GND plot for M31}

The lower panel of 
Fig.~\ref{MWM31dp} reveals,  as in the MW,
 two 4-galaxy-normal   density peaks in M31 (reference satellite sample M31\_34, see Sec.~\ref{sec:obsdata}).
In this case they both show  comparable strengths (see Table~\ref{c1mwm31}) and are located quite separate
% ($\sim80^{\circ}$) 
from one another.
The peak strength values are lower than those of the MW. This is in part due to the higher M31 satellite distance errors (as compared to those of the MW), which blur the peaks in the 4GND plot.

We define Peak 1 
as the over-density at
% $(l,b)=(26.60, -6.14)$,
  $(l,b)=(26.6^\circ, -6.1^\circ)$,
 and Peak 2 as that located at  
% $(l,b)=(-69.14 , 21.68)$. 
  $(l,b)=(-69.1^\circ  , 21.7^\circ)$.
The direction of
 M31's spin vector, depicted with an 'X', indicates that
the planar configurations defined by both peaks
  are not perpendicular to the galaxy's disc, but inclined 
%$48.93^\circ$ and $70.49^\circ$, respectively.
$\sim49^\circ$ and $\sim70^\circ$, respectively.
Interestingly,
Peak 1 forms an  angle of 
%$83.86^\circ$
$\sim84^\circ$
 with the MW's spin vector, meaning its corresponding planes are approximately normal to the MW's disc
 \citep{Conn13}. 
Furthermore, the projected angular distance on the sphere between Peak 1 and Peak 2 is of
 %$82.43^{\circ}$,
  $\sim82^{\circ}$,
 and the angle between Peak 1 (Peak 2)  and the Sun - M31 line is 
 %$87.73^{\circ}$ ($5.31^{\circ}$).
  $\sim88^{\circ}$ ($\sim5^{\circ}$).
  Therefore the planar configuration of satellites defined by Peak 1  
  is observed nearly edge-on from the MW
(see table 5 in P13),
  % (see also \citealt{Pawlowski13} table 5), 
while that of Peak 2   is approximately perpendicular to it and would be observed mostly face-on.

%parrafo q habia aqui lo muevo de aqui al final d la subseccion

Fig.~\ref{barM31} shows that approximately half of the satellite sample contributes dominantly to each corresponding peak $p$.
Focusing on the identities of satellites, we find that,
out of the 16 satellites with highest $C_{1, s}$, 9 are among the satellites with lowest $C_{2, s}$.
On the other hand,
out of the 16 satellites with highest $C_{2, s}$, 9 are among those with lowest $C_{1, s}$.
This is indicating that the two over-densities' contributing members are not the same.
While  satellites contributing most to Peak 1 define the GPoA plane from 
P13
%\citet{Pawlowski13}
(and also generally coincide with satellites in ``plane 1" from \citealt{Shaya13}),
Peak 2 defines a separate and independent predominant planar satellite configuration.
%%%Peak 2 and its corresponding predominant planar satellite configuration 
This structure corresponding to  M31\_34 Peak 2 
has not been analyzed prior to this study\footnote{Note that the  planar structure derived here from Peak 2 does not correspond to the so-called 'M31 disc plane' noted in \citet{Pawlowski13}, or to ``plane 2"  in \citet{Shaya13}. }
and will be described in detail below.

We note that an analysis of the M31\_36 sample (i.e., including the satellites AXVI and AXXXIII too) gives  quite similar results. As an illustration, in Table~\ref{c1mwm31} we give  the values of the corresponding $C_1$ and $C_2$ peak strengths. The $C_2$ values of M31\_34 and M31\_36 are consistent with each other within their error bars, while the  $C_1$ strengths are close to be consistent too, with Peak 1 in M31\_36 slightly stronger  than its M31\_34 counterpart.
Indeed, for M31\_36 we find the same plane structure  as in M31\_34; i.e., 2 independent peaks pointing in  the same normal directions as quoted above for M31\_34. In the M31\_36 case, AXVI turns out to be the main contributor to Peak 1 (explaining the slight increase of $C_1$ strength), while satellite AXXXIII does not contribute to Peak 1, and only mildly to Peak 2.

%parrafo al final de la seccion M31

\subsection{Quality analysis}

\begin{figure*}
\centering
\caption{ Edge-on view of 
 M31\_34-2-18, the highest-quality plane from Peak 2 in M31\_34   with $N_{\rm sat}$=18 members (in red),
and  the GPoA (in blue), showing their relative orientation.
M31's galactic disc and spin vector are depicted in green. 
Satellites belonging only to the M31\_34-2-18 plane are shown as red points, while satellites belonging only to the GPoA are shown as blue points. Satellites shared by both samples are violet.
}
\includegraphics[scale=0.9]{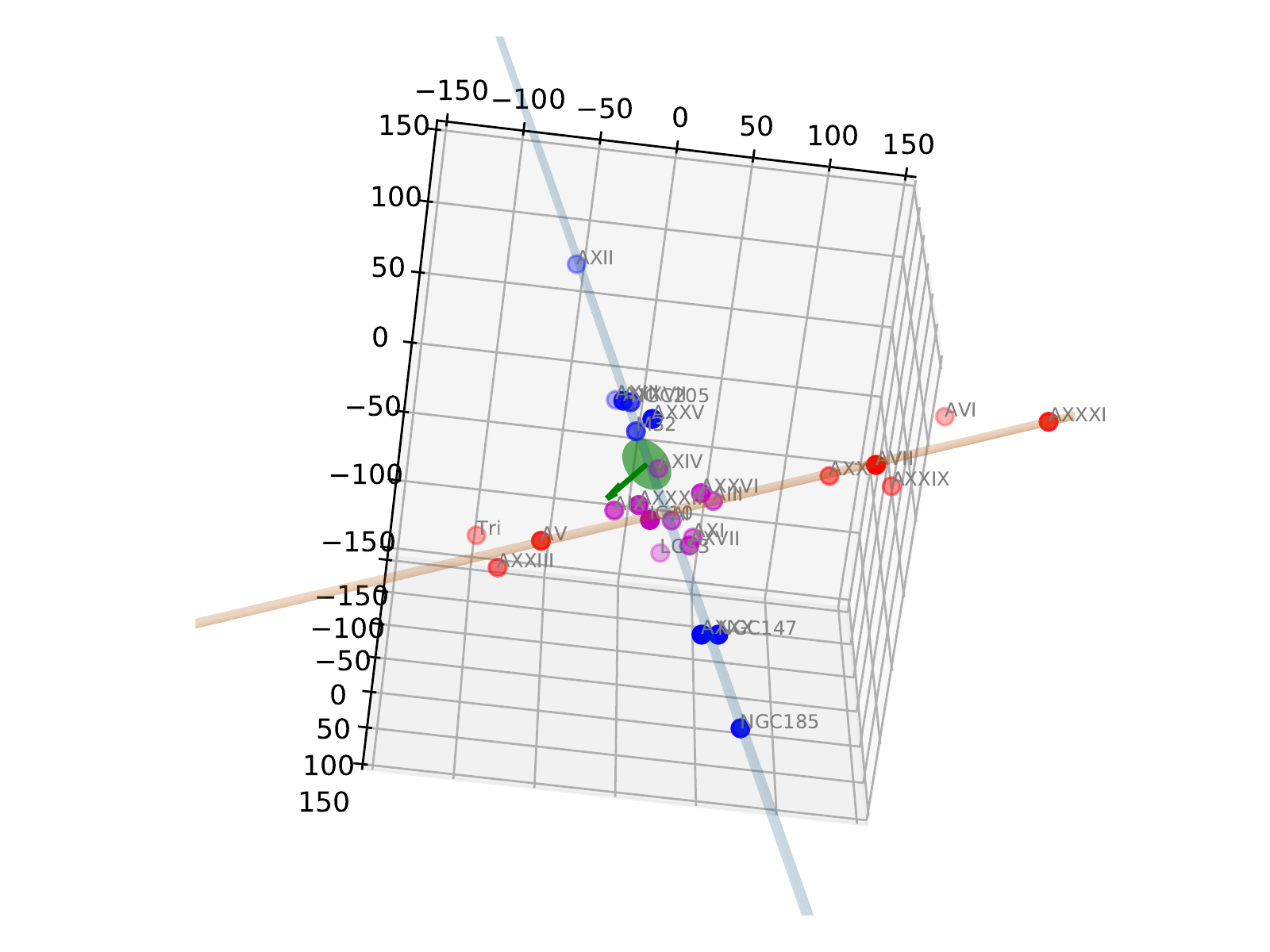}
\label{3D_M31_p1p2}
\end{figure*}

For each peak, 
we build up their respective collections of planes of satellites by
 iteratively applying the best-fitting 
 plane technique to an increasing number of satellites  $N_{\rm sat}$,
following the order given by Fig.~\ref{barM31}.
The 
%values of concentration ellipsoid parameters 
parameters resulting from the ToI fitting
versus $N_{\rm sat}$ are shown in Fig.~\ref{M31_ca_frac}.
We see that  $c/a$ and  $b/a$ take respectively low and high values, confirming that these spatial distributions are actually planes up to $N_{\rm sat} \sim 24 - 25$. When considering the whole sample of M31\_34 satellites, $c/a$ and $b/a$ take a similar value of $\sim0.6$, indicating instead a non-flattened ellipsoidal spatial distribution
(as two approximately perpendicular planes cross).

The corresponding parameter error bands are clearly apparent in the case of M31 as compared to the MW, due to overall larger satellite distance uncertainties.
As mentioned previously, 
while the GPoA is viewed approximately edge-on from the Sun, the planes of satellites from Peak 2 
 are viewed mostly face-on. Therefore the uncertainties in
the Sun - satellite distances affect more (less) to $c/a$ 
%%and $\Delta$RMS parameters
 than to $b/a$, in the case of Peak 2 (Peak 1). 
This fact explains the different magnitude of the error bands of  ToI parameters
corresponding to different peaks
 shown in this figure.

Focusing on the $c/a$ 
%%% and $\Delta$RMS 
panel, one can see that only up to $\sim$half 
of the total number of satellites contributing  respectively to Peak 1 and Peak 2
form a thin planar structure, which rapidly thickens  as more members are added to the plane-fitting iteration.
Moreover, as said above, satellite identities contributing most to both peaks are overall different, as shown in \Fig{barM31}.
Therefore,  in contrast to the MW27 and MW46, the M31\_34 satellite  sample does not form one preferential planar structure but seems to be divided in 
%(at least) 
two.

The normal directions to the planes are  stable as $N_{\rm sat}$ increases and reaches  $N_{\rm sat}=$19
for Peak 1 and $N_{\rm sat}$=18 for Peak 2, showing that the two satellite planar structures are well defined. 
Beyond these values, the normals to the corresponding planes are not that well fixed.

As for the distances, planes around Peak 1 pass very close to the M31 center, while planes belonging to Peak 2 do not
(see Fig.~\ref{M31_ca_frac}).
The values of $D_{\rm cg}$  in this case range between $\sim12-34$ kpc,
within, or close to, the values of $\Delta$RMS for planes associated to Peak 2 (see Table~\ref{table_obsThisWork}). However, the question arises if, given 
%, and specifically,the M31\_34-2-18 plane presents $34.9\pm11.2$ kpc.  
  the large radial distance errors 
and the  uncertain potential well of a complex, binary system as is the Local Group, 
%it is unclear
 if these distance values are still within reasonable ranges to allow for  the possibility of dynamical stability.

The specific   values for the  M31 Ibata-Conn-14 and GPoA planes
are shown with
colored circles in Fig.~\ref{M31_ca_frac} (see \Tab{table_obspl}).
 Our methodology reveals a combination of $N_{\rm sat}$= 14 satellites (marked with a black open circle) that yields a higher quality plane than the `Ibata-Conn-14' plane
(see M31\_34-1-14 entry in  \Tab{table_obsThisWork} and Fig.~\ref{barM31} for satellite identities).  
  This occurs because the  `Ibata-Conn-14' plane was defined among only PAndAS survey satellites,
while the sample used 
in 
P13
%\citet{Pawlowski13} 
and 
here includes
the PAndAS satellites 
within 300 kpc of M31 (25 out  of 27)
plus 
9  satellites discovered differently
(i.e., LGS3, IC10, AXXXII, AVII, AXXIX, AXXXI, AVI, NGC205, M32).
Interestingly, the latter turn out to be   precisely among the dominant 4-galaxy-normal contributers to both M31\_34 Peaks 1 and 2.

In turn, the magenta circles  corresponding to the GPoA match the solid line (Peak 1)
because the 19 satellites that we find with highest $C_{1,s}$ are precisely  the GPoA satellite sample.  
%\red{Note that we find no correlation between $M_{\rm star}$ and $C_{1s}$ at a higher than 76\% confidence level,
%(log10, log10)      -0.05553313082435409    0.7550862006457574
%and  between $M_{\rm star}$ and $C_{2s}$ at a higher than 65\% confidence level.}
%(log10, log10)     0.08165028349571857      0.6461921169440874
Note again that the satellite system shows low correlation coefficients $r$  between 
$M_{\rm star}$ and $C_{1,s}$ or $C_{2,s}$,
with a  higher than 75\%  probability of getting such $r$ values assuming that there is no correlation (see  Fig.~\ref{cps_mstar}).

Moreover, our analysis allows the identification of the highest quality planes in M31 as  $N_{\rm sat}$ increases at $c/a$ roughly constant.
These planes correspond to the points in Fig.~\ref{M31_ca_frac} at which the M31\_34 Peak 1 and Peak 2 lines start to
increase rapidly  in the $c/a$  panel,
 marked in Fig.~\ref{M31_ca_frac} with
black open circles.
 For Peak 1 
we single out a good quality plane at low $c/a$
 %%the highest quality plane  at low $c/a$ occurs
   with $N_{\rm sat}$=16 (M31\_34-1-16  in Table~\ref{table_obsThisWork}).
 %, with   $c/a$=0.06, $b/a$=0.56 and  $\Delta$RMS= 7.48 kpc.  
 %
We also note the high quality of the plane with $N_{\rm sat}$=23  (M31\_34-1-23), with 
  %$c/a$=0.18, $b/a$=0.72 and  $\Delta$RMS= 20.98 kpc. These
 ToI parameter values  very similar to
  those of the VPOS-3 plane of satellites  in the MW.

For Peak 2,
we highlight the plane
with $N_{\rm sat}$=18 members (M31\_34-2-18), 
beyond which the normal vector $\vec{n}(l,b)$ to the best-fitting plane does not conserve its direction.
%, yielding a planar distribution with  $c/a$=0.16, $b/a$=0.73 and $\Delta$RMS= 20.45 kpc parameters.
%This  high quality plane from Peak 2 
%The latter
This plane
 presents  comparable properties to the GPoA (magenta circle),
 to which it is roughly perpendicular. 
Given  that the GPoA has one satellite more and a lower $c/a$ value than the M31\_34-2-18 plane (see Tables~\ref{table_obspl} and \ref{table_obsThisWork}), strictly speaking the former
 has a higher quality than the latter. However, the differences are a 5\% in $N_{\rm sat}$, and a 10\% in the $c/a$ values,
if we take into account the error bars. 
Therefore we can conclude that the qualities of both planes are comparable.

Fig.~\ref{3D_M31_p1p2} shows the relative orientation 
between M31\_34-2-18 and the GPoA
in a coordinate system oriented such that both planes are seen edge-on.
Note that 10  satellites are shared by both samples (i.e., LGS3, IC10, AXIV, AXI, AXXXII, AI, AXVII, AIX, AIII and AXXVI, in violet in the figure).

%% It is of interest whether this plane could be dynamically stable, this is, if the member satellites corotate within the planar structure they define in space. Line-of-sight velocities of the M31-2-18 satellites \citep[taken from][]{McConnachie12,Collins13,Martin14}, which we observe face-on from the MW, give a perpendicular velocity dispersion of $\sigma=90.20$ km/s. 
%% According to \citet{Fernando17}, such a plane will be erased in a short timescale and is just a fortuitous alignment of satellites, as they find that 
%% planes with a perpendicular velocity dispersion above $\sim50$ km/s disperse to contain half their initial number of satellites in 2 Gyrs time.

\begin{figure}
\centering
\includegraphics[width=\linewidth]{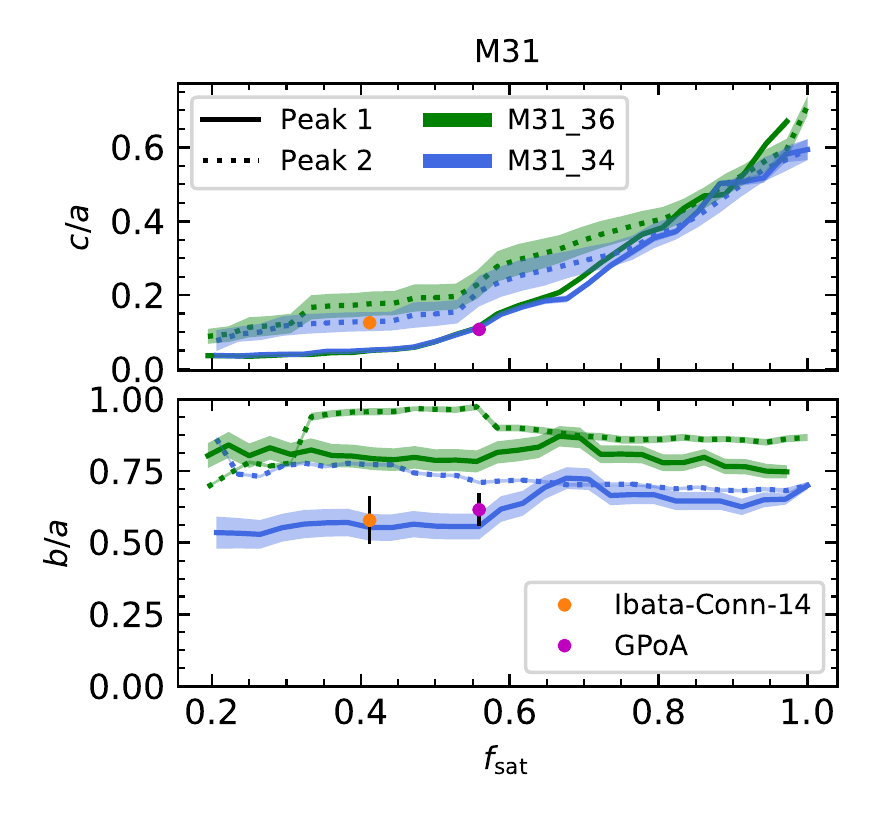}\\
\vspace{-0.6cm}
\caption{The $c/a$ and $b/a$ ratios in terms of the satellite fraction $f_{\rm sat}$. Blue (green) lines are results for the M31\_34 (M31\_36) satellite sample.
}
\label{M31-3436_fsat}
\end{figure}

The quality analysis of the planar structures arising from the M31\_36 sample returns similar results to those just presented for M31\_34. 
In most cases,  the resulting  ToI parameters are consistent with those of M31\_34 within 1$\sigma$, or close to being consistent. 
The only remarkable difference lies in  the values of the $b/a$ ratios, larger for the M31\_36 planes  than for the M31\_34 ones  with same $N_{\rm sat}$. This occurs because the structure expands as two further away satellites are added to the sample.
Fig.~\ref{M31-3436_fsat} illustrates this behaviour  in  terms of the satellite fraction.  
Changes of slopes in the $c/a$ versus  $N_{\rm sat}$ plot indicate that
the high-quality planes to be singled-out  have now 1  more satellite for Peak 1, (i.e., M31\_36-1-17; M31\_36-1-24), or 2 more satellites for Peak 2 (i.e., M31\_36-2-20) as compared to those singled out in the  M31\_34 sample.
%, although with  no qualitative changes.    SI  lo crees necesario / conveniente se puede dar en una Tabla los valores de los parameros TOI, ya calculados. 
Their specific ToI values are given in Table~\ref{table_obsThisWork}.

We note that the previously described results for M31 are based on the heliocentric destance moduli given in Table~\ref{tab:m31data}, that are those provided in \citet{McConnachie12}'s data compilation. Using instead distance data from table 2 in \citet{Conn2012} based solely on the ground-based TRGB does  not change the results shown in Fig.~\ref{M31_ca_frac} significantly, and therefore they will be not be made explicit here.

It is of order, nonetheless, to emphasize the sensibility of M31's Peak 2 planar structure to satellite heliocentric distances,  due to its face-on orientation relative to the line-of-sight from the Sun.
%Indeed, M31 satellites are  currently endowed with large distance uncertainties independently of the method (TRGB, HB, RR Lyrae, etc.) used \citep[see discussions in][]{MartinezVazquez2017,Weisz2019}. 
%
On one hand, distance uncertainties are indeed large, but these are taken into account and the Peak 2 planar structure is robust in our analysis.
However, 
 offsets between the \textit{mean} distance moduli for the same object
coming from different methodologies
\citep[HST- or ground-based TRGB and HB; RR Lyrae, etc; see e.g.][]{McConnachie05,Conn2012,MartinezVazquez2017,Weisz2019}
 can be in some cases 
  larger than a few times the average  thickness of the Peak 2 plane.
This is  due to calibration issues in each methodology (see discussions in the previous references).
Thus, 
 any changes to the distance data used, resulting in a distance dispersion
relative to the data assumed here (Table~\ref{tab:m31data})
 of the order of a few times the  $\Delta$RMS of the plane  ($\sim 20$ kpc, see Table~\ref{table_obsThisWork}), could end  up possibly erasing the M31 Peak 2 planar structure (while keeping that corresponding to Peak 1, whose orientation is  close to edge-on with respect to the Sun).
The robustness of the Peak 2 plane of satellites will thus have to be confirmed in the future   as more accurate
and homogeneously-measured
 M31 satellite distances are available.

%%%%%:::::::::::::: M31-34 and M31-36  TOGETHER     ::::::::::::::::::

%%%:::::::::::::  NEW  SECTION :::::::::::::::

\section{Using full six-dimensional phase-space information}\label{sec6D}

%%%::::::::::::  CO-ORBITATION TEST :::::::::::::::

\subsection{Co-orbitation within  the positionally detected high-quality planes}
\label{co-orbitation}

Proper motion data for MW satellites has revealed one important feature of the plane of satellites observed in the MW: that it presents a high degree of coherent rotation  \citep{Metz08,Fritz18,Pawlowski2020}. Indeed, the derived 
orbital angular momentum vectors (i.e., orbital poles, $\vec{J}_{\rm orb}$) of several MW satellites
%%the constituent satellites 
 are aligned with the normal to the VPOS-3 within a small angular distance. In particular, \citet{Fritz18}, 
define co-orbiting satellites within a given plane 
%are taken  
as  those whose orbital poles lie within an angular distance of $36.87^\circ$  around the  normal to the plane.
This corresponds to an area of 10\% of the sphere surface (or 20\%,  when not differentiating objects co-rotating or counter-rotating with the disc of the central galaxy).

We study if the MW satellites included in the high-quality planes detected with the 4GND plot method 
 are co-orbiting within them. To this end we measure the angular distance between the normal vector to a given singled-out plane, $\vec{n}$, and each individual satellite orbital pole  $\vec{J}_{\rm orb}$. 
We consider 36 satellites from the MW46 sample,  for which kinematic data is currently  available. In particular, the orbital poles for the LMC and SMC are taken from \cite{GaiaHelmi18}, while those of the rest of satellites are taken from \cite{Fritz18}, \cite{Fritz2019} and \cite{Torrealba2019}.\footnote{No proper motions and/or radial velocities have been measured yet for  Pictor2, Virgo1, Bootes4, PegasusIII, Centaurus1, Cetus3, Tucana4 and Grus2. Moreover, proper motion data for Columbus1 and Reticulum3 as measured by \cite{Fritz2019} and \cite{Pace2019} are very different with each other and endowed with larger errors than for the rest, therefore we prefer not to consider them for our kinematic analysis here. 
} 
%%%We also take from the latter reference their 2000 Monte Carlo simulations of orbital poles including measurement errors.
 We refer the reader to these papers for details.\footnote{ 
We note that this is indeed a rapidly evolving field.  \cite{Fritz2019} and \cite{Torrealba2019}  are recent measurements that have appeared during the revisions of this manuscript. Also after  the finalising of this paper, improved  proper motions for MW satellites by \citet{McConnachie2020} have become public. The analyses presented here will thus have to be updated as new data becomes available (e.g., with \textit{Gaia}-DR3).}

The projection on the sky of these orbital poles is shown in green in Fig.~\ref{3jorb}, with the actual direct measurements depicted as circles with labels, while 2000 Monte Carlo simulations of the poles including measurement errors, are shown as tiny points.
In this Aitoff diagram, orbital poles are shown as all co-rotating with the disc of the MW. Objects that are actually counter-rotating are distinguished by a lighter green color. 
Note however that,
for those MW satellites whose orbits are close to  perpendicular to the Galactic disc,
 the fact that they are measured as rotating or counter-rotating is not well-determined, but within the errors.

\begin{figure}
\centering
\vspace{-0.5cm}
\includegraphics[width=\linewidth]{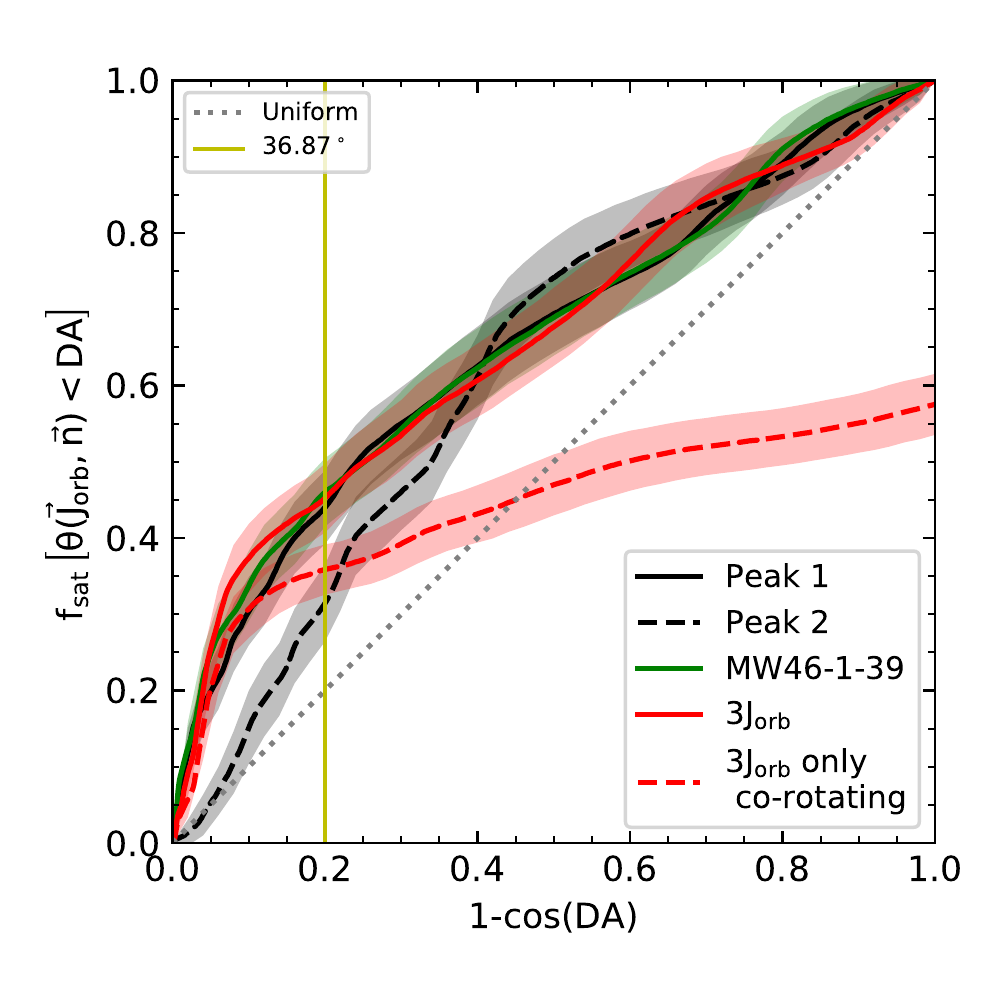}
\vspace{-0.8cm}
\caption{ Fraction of MW46 satellites with orbital poles enclosed in a given angle $DA$ measured from  different reference axes: 4GND plot Peaks 1 and 2 (black solid and dashed line, respectively); the normal to the MW46-1-39 plane (green), and  the main peak of the 3J$_{\rm orb}$-barycenter density plot (red).
We do not differentiate between co-rotating or counter-rotating orbital poles with the disc of the galaxy. Only for the last axis do we show as well the results of only co-rotating MW satellites (red dashed).
Solid lines show the median values while shaded areas the $\pm\sigma$ range. Uncertainties have been computed from assuming 1000 random realizations of the orbital poles including measurement errors.
% (Monte Carlo simulations from \citet{Fritz18}). 
 The dotted line shows the result for a uniform distribution, and a yellow vertical line marks an angle of 36.87$^\circ$.   
In \citet{Fritz18}, MW satellites with orbital poles enclosed within this angle from the VPOS-3 are considered to be co-orbiting within the plane. 
}
\label{cosA_MW}
\end{figure}

Fig.~\ref{cosA_MW} shows the
fraction of the total number of satellites with  $\vec{J}_{\rm orb}$ enclosed by a certain angle $DA$  from the reference axis $\vec{n}$
(note that in this figure we do not differentiate  between co-rotation or counter-rotation with the  disc of the central galaxy, therefore the angular distance $DA$ can be a maximum of 90$^\circ$).
Solid and dashed black lines  give the clustering results around the directions of MW46 Peak 1 and 2, respectively. These lines show the mean value of 1000 Monte Carlo random realizations of orbital pole projections,
while a shaded area shows the $\pm 1\sigma$ dispersion range.
A yellow vertical line is drawn at $DA=36.87^\circ$, while a dotted line shows the expected result for an isotropical distribution of orbital poles.
It is clear that the normal direction to Peak 1 defines a plane with a higher degree of satellite co-orbitation, than that of Peak 2. In particular, there is a
$44\pm5\%$ of co-orbiting satellites around Peak 1 and a $31\pm5\%$ around Peak 2.
 %%%%$40.6^{+6.3}_{-3.1}\%$ of co-orbiting satellites around Peak 1 and $28.1^{+6.3}_{-3.1}\%$ around Peak 2.
Note that if assuming directly the measured orbital poles (circles in Fig.~\ref{3jorb}), these fractions increase to
$50\%$ and $39\%$,
 %%$46.9\%$ and $37.5\%$,
  respectively.

When using instead as reference axis the normal directions to specific singled-out planes from MW46   
we obtain a very similar result to that with the Peak 1 axis. As an example, we show with a green line the clustering obtained with respect to the normal vector to
MW46-1-39 (the normal directions to Peak 1 and MW46-1-39 are separated only by $2.8^{\circ}$).
%% MW37-1-32 (the normal directions to Peak 1 and MW37-1-32 are separated only by 4.9$^\circ$).

%%%The red line has as reference axis the main clustering direction detected, for the same MW37 subsample, with the \textit{3Jorb-barycenter method}, see next subsection. 

These results indicate that
approximately half 
 %%%an important fraction 
 of the satellites belonging to positionally-detected, high-quality planes
 share coherent motion, while the other half 
could mostly
 %%% are interlopers, and will 
  be lost to the plane in a short time after observation. 
   Results qualitatively consistent with these ones have also been found in cosmological simulations 
\citep[see Paper II,][]{Gillet15,Buck16,Shao19}.

Concerning M31, it is of interest whether the  M31\_34-2-18 plane could be dynamically stable, this is, if the member satellites co-orbit within the planar structure they define in space. 
While in the lack of proper motion measurements for M31 satellites, 
line-of-sight velocities of the satellites involved  in the M31\_34-2-18 plane \citep[taken from][]{McConnachie12,Collins13,Martin14}, which we observe face-on from the MW, give a perpendicular velocity dispersion of $\sigma=90.20$ km/s.
 According to \citet{Fernando17}, such a plane will be erased in a short timescale and is just a fortuitous alignment of satellites, as they find that
planes with a perpendicular velocity dispersion above $\sim50$ km/s disperse to contain half their initial number of satellites in 2 Gyrs time.

This result should be taken with caution, and confirmed when 3D-velocities are available, as numerical experiments  have shown that line-of-sight velocities are not well representative of the former \citep{Buck16}. We note even so, that, as this plane is observed face-on from the MW, it is not expected that the consideration of the tangential velocities may change much the perpendicular velocity dispersion estimated here.
What may be more relevant to this face-on plane, however, are the very uncertain radial distance measurements to M31 satellites, as discussed earlier.
% As discussed earlier, offsets between the distance moduli assuming different methodologies can be larger than the average thickness of the plane.  
Thus, until this position data achieves higher accuracy and consistency between methodologies, 
the  outcomes of a dynamical assessment of this spatial structure will remain less convincing.

%\green{We note, however, that  \textbf{this should be taken with caution as numerical experiments  have shown that }
%the reconstruction of the three-dimensional velocity field from line-of-sight velocities  
%\textbf{ is very sensitive to  the viewing angle value,  providing numbers of co-rotating satellites that
%are consistent with those obtained from a sample with random velocities
%\citep{Buck16}. } }

%%%::::::::::::::  DETECTION ::::::::

\subsection{Identifying the axes of maximum satellite co-orbitation}
\label{3JORB_method}

Neither the 4GND plot method, or its extension, are designed to identify planes of kinematically-coherent satellites able to persist in time, as they focus only on the position of satellites.
In Paper III (Santos-Santos, in preparation)
the 3J$_{\rm orb}$-barycenter
 %%a 
 method is introduced,
which leads to the identification of axes around which a maximum number of satellites co-orbit out of a sample of $N_{\rm tot}$ of them.  Combining the results at different times,  high quality, persistent planes of satellites are identified in simulations of disc galaxies.
 %%% to \textbf{the identification of} high quality, persistent planes \textbf{in simulations of disc galaxies}.

Briefly, this method  consists in finding clustering in the
space of orbital angular momentum vector projections,
 %%orbital \textbf{angular momentum} poles  space, 
 following a protocol similar to that of the 4GND plot method  \citep[see also][]{Garaldi2018}. 
 Given the orbital poles of a satellite set, the barycenters of all the spherical triangles\footnote{We discard all spherical triangles with interior angles larger than 60$^\circ$ to avoid spurious overdensities at the poles.} that can be formed out of the 
$N_{\rm tot}$ poles are calculated. This gives $N_{\rm 3Jorb}$ points on the sphere (the combinations of $N_{\rm tot}$ over  3).
Such a method multiplies the statistics as an asset to search for clustering within an otherwise small set of satellite orbital poles.
 A smoothing  similar to that described in Section~\ref{4gnd_method}, leads to the so-called 3J$_{\rm orb}$-barycenter density plot, whose maxima represent directions around which the orbital poles cluster. The method allows us to determine where these maxima are placed  on the sphere; that is, the 
normal directions to 
% poles of 
 planes that are kinematically-coherent by construction.

\begin{figure*}
\centering
\includegraphics[width=\linewidth]{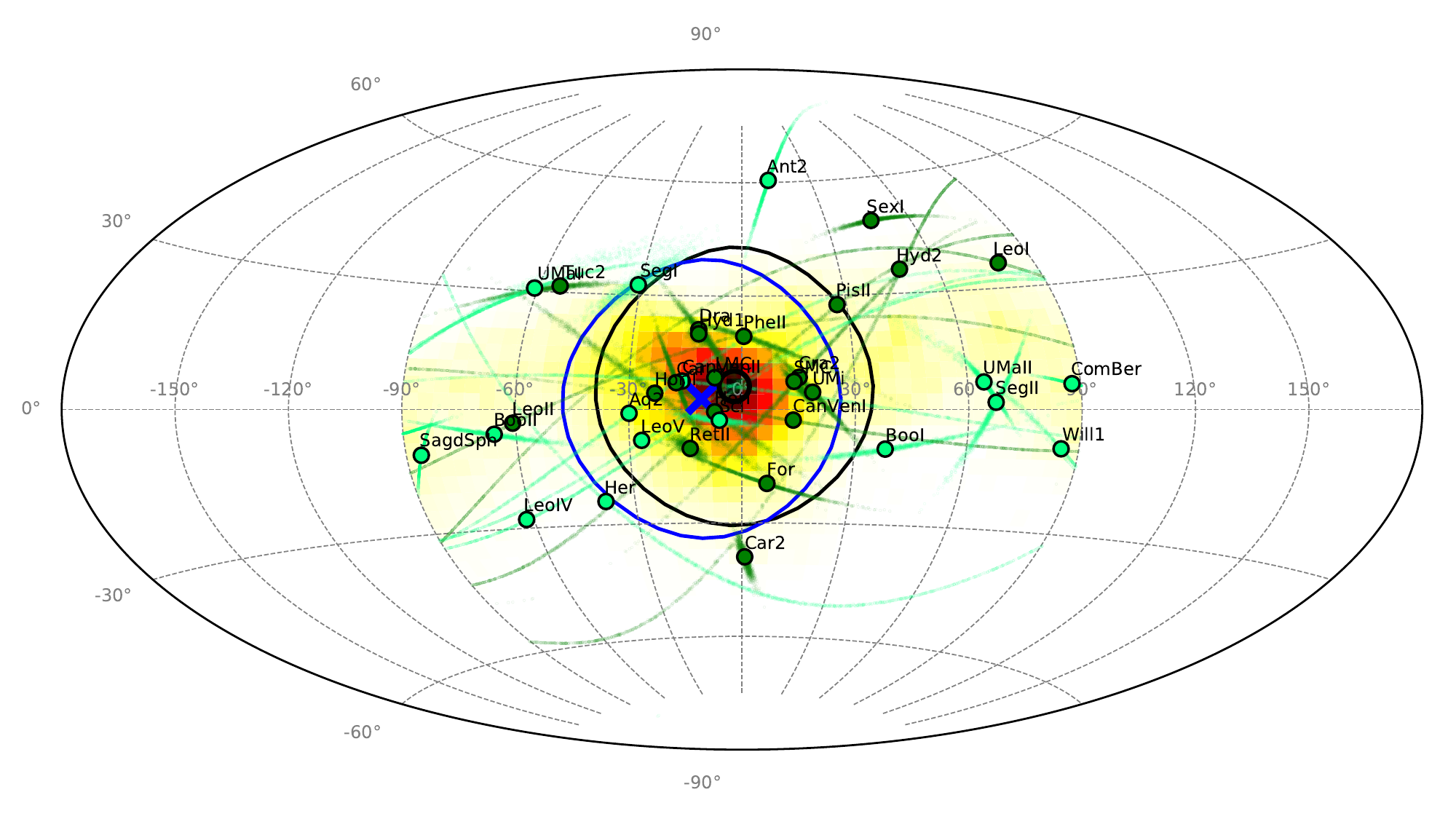}
\caption{
Aitoff diagram showing the orbital poles of MW satellites \citep[data from][]{GaiaHelmi18,Fritz18,Fritz2019,Torrealba2019}. Actual measurement values are shown as circles and 2000 Monte Carlo simulations including uncertainties are shown as clouds of small points. 
These are Galactocentric coordinates, centered in the Galactic \textit{center}. All orbital poles are plotted within [-90$^\circ$,+90$^\circ$] in longitude, as if they all co-rotate with the MW's Galactic disc. 
A lighter green color marks satellites that are actually counter-rotating. 
%A darker green color marks satellites that are co-rotating with the MW's Galactic disc.
In the background is the density plot resultant of applying the '3Jorb-barycenter method' (See Section~\ref{3JORB_method}).
% and Paper III in prep.). 
It has been computed by stacking the result of 100 random realizations of the orbital poles, taking into account measurement errors. The densest bin, and therefore the direction of the main co-orbitation axis found with this method, is circled in black.
A blue cross marks the normal direction to the VPOS-3 plane. 
Additionally, we show circles withstanding an angle of 36.87$^\circ$ from the previous axes (see text). 
%This corresponds to an area of 10\% of the sphere. 
%In \citet{Fritz18}, satellites with orbital poles enclosed within this angle from the VPOS are considered to be co-orbiting within the plane. 
%%The legend shows the total number of '3Jorb-barycenters' used to compute the density plot.
}
\label{3jorb}
\end{figure*}

This method has been applied to the subsample of MW46 satellites for which kinematic data are available, described in the previous subsection.
The corresponding 3J$_{\rm orb}$-barycenter density plot is drawn in Fig.~\ref{3jorb}, in the background of the Aitoff diagram.  In particular, 100 realizations have been stacked, where we consider 100 orbital pole directions taken at random from the 2000 Monte Carlo simulations
% calculated in \citet{Fritz18}
 including the measurement errors.
One unique remarkable overdensity stands out, with its peak at 
%$(l,b)=(-2.05, 6.14)$ 
$(l,b)=(-2.1^\circ, 6.1^\circ)$ 
(the densest bin is circled in black). 
This axis is 
only 4.6$^\circ$  apart from
%%very close to 
the normal direction to the  VPOS-3, marked with a blue 'X' (see Table~\ref{table_obspl}), and therefore also very close to  the normals of all planes corresponding to Peak 1 found with the  position-only 4GND plot method (see Table~\ref{table_obsThisWork}).

We use now the new axis found with the 3J$_{\rm orb}$-barycenter method to compute the fraction of co-orbiting MW satellites. This is shown in red in  Fig.~\ref{cosA_MW}. 
There is a slightly higher fraction 
%(\textbf{XXX\%})
 of co-orbiting satellites 
at small $DA$ 
 %within 36.87$^\circ$ 
 than in the cases when assuming  the normal directions to the positionally-detected planes (black lines),
 but 
approximately 
 the same mean and dispersion is found at $DA=36.87^\circ$ than for Peak 1.
 Translating to absolute numbers, we find  $16\pm2$ co-orbiting satellites  out of the considered sample with $N_{\rm tot}=36$. 
 If focusing on the actual measured values, $18$ satellites co-orbit (see objects within big black circle in Fig.~\ref{3jorb}, i.e., PiscesII, Draco, Hydrus1, Carina, CanesVenaticiII, LMC, SMC, Crater2, UrsaMinor, CanesVenaticiI, Aquarius2, Carina3, Sculptor, LeoV, Fornax, ReticulumII, HorologiumI, PhoenixII).
Interestingly, if only considering co-rotation with the Galactic disc,  the mean fraction of co-orbiting satellites is $36\%$, which corresponds to $13$ satellites ($14$ if assuming the actual measurements), which shows that counter-rotating satellites in this structure are a minority (only $4$, i.e., CanesVenaticiII, Aquarius2, Sculptor, LeoV).

\begin{figure}
\centering
\includegraphics[width=\linewidth]{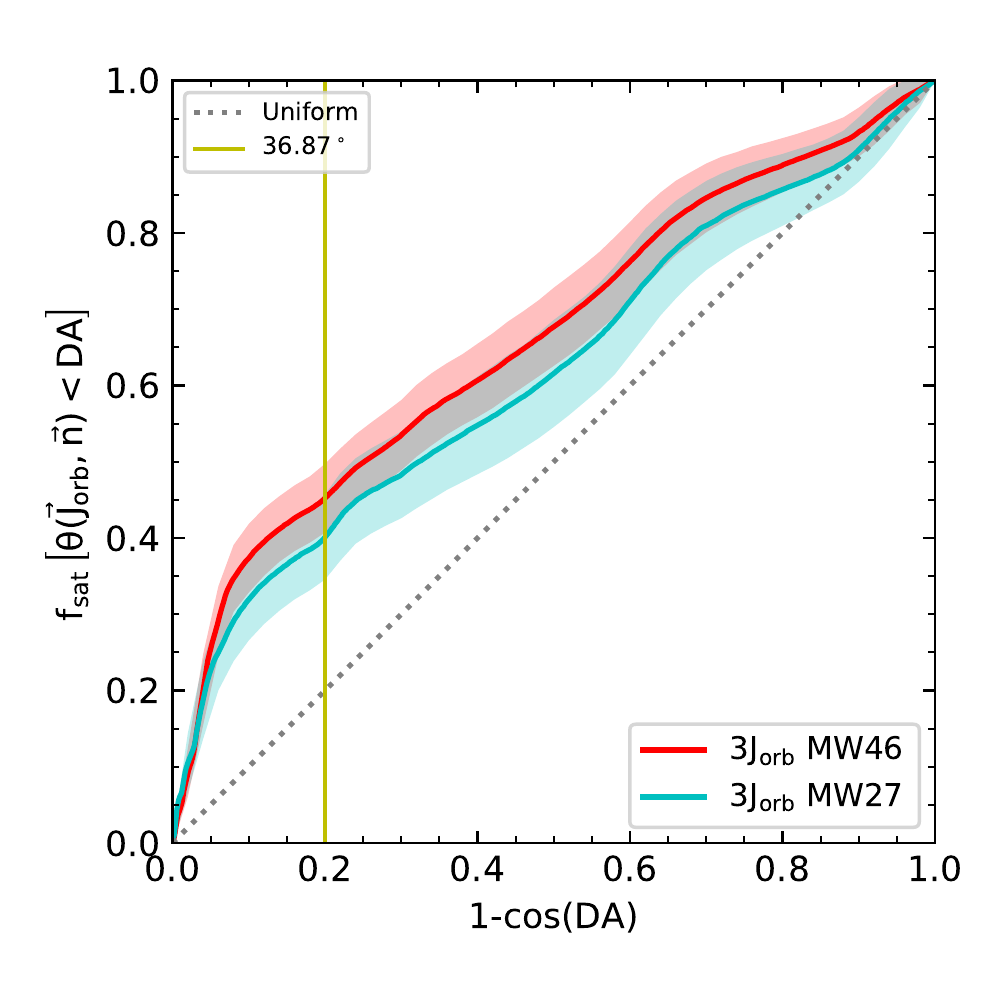}
\caption{
Same as Fig.\ref{cosA_MW}, comparing the results  of the 3J$_{\rm orb}$-barycenter method applied to the samples of MW
satellites with kinematical information drawn from MW27 and MW46 (see Table~\ref{tab:mwdata}).  }
\label{unocos_27}
\end{figure}

  For the sake of comparison, the MW27 sample has also been kinematically analyzed through the 3J$_{\rm orb}$-barycenter method. In this case, 25 of its members have kinematic information. 
Again, a unique remarkable 3J$_{\rm orb}$ axis has been identified, which happens to be very close to that of the MW46 satellites 
%$(l,b)= (-1.02, 5.11)$.
$(l,b)= (-1.0^\circ, 5.1^\circ)$.
The fraction of the 25 satellites with poles at angular distance $<DA$  around this axis is given in Fig.~\ref{unocos_27}, where we also plot the corresponding result for the MW46 satellites (red solid line in Fig.\ref{cosA_MW}). They show 
indistinguishable clustering properties within the error bars, with a slighly larger mean in the case of MW46. Therefore, with the current kinematic data and within the error bands, the newly identified satellites relative 
to MW27 follow the kinematic  structure already present with MW27 satellites.

These results show how this new method directly identifies a rotation axis whose  corresponding normal plane contains the maximum possible number of satellites co-orbiting within it.
%\textbf{Indeed, with this method it is possible to identify %%this method allows the identification of 
%the satellites constituting} the largest set of kinematically-coherent satellites that can be formed out of \textbf{the total $N_{\rm tot}$.}
In paper III the method is applied to two zoom-in hydro-simulations of disc galaxies, and a persistent plane,
with the same satellite members along cosmic evolution,
 %of kinematically-coherent satellites 
 is identified in each case.

%%%:::::::::::::::::::::::::::::::::::::::::::

\section{Summary and conclusions}\label{conclu}

Recent studies on planes of satellites have resorted to evermore refined methods to define and characterize them. 
In this work, we have further developed one such method,  the '4-galaxy-normal density plots' method
(P13)
with an extension designed to 
identify, systematically catalog and 
  study in detail the \textit{quality} of the predominant planar configurations 
 revealed by  over-densities in the  4-galaxy-normal density plots.
The method has been applied to
the currently considered confirmed satellite galaxies of the MW and M31 systems
 %up-to-date data of the MW and M31 confirmed satellite systems 
 (no nearby dwarfs, globular clusters or streams have been considered), 
providing the most detailed and updated characterization of their satellite planar-like spatial distributions. These serve as 
 %providing 
 a reference with which to compare results from numerical
hydrodynamical simulations of disc galaxy systems
(see, e.g., Paper II).

We count the weighted number of times each satellite $s$ contributes to 4-galaxy-normals within 15$^\circ$ of a  specific over-density $p$
identified in a density plot
 ($C_{p,s}$).
For each relevant over-density, 
we define its peak strength as $C_p = \sum_s C_{p,s}$, normalized to the total number of 4-galaxy-normals.
We order satellites   by decreasing $C_{p,s}$ and 
iteratively fit planes to subsamples of satellites following this order.
In this way, rather than a plane per over-density,
we yield a catalog or \textit{collection of planes} of satellites  with an increasing number of members $N_{\rm sat}$,
 whose normals cluster around the density peak. 
% \textbf{The low $N_{\rm sat}$ planes selected by the method  are part of the predominant planar structure consisting of a high $N_{\rm sat}$, signaled by the over-density.}
This method selects the subsamples of $N_{\rm sat}$ satellites with lowest $c/a$ provided that they are embedded in a larger, populated, predominant planar structure. Indeed, in the case of planes with high $N_{\rm sat}$/$N_{\rm tot}$, the method selects the subsample of satellites that yields the thinnest plane out of all possible combinations of $N_{\rm sat}$ satellites.  On the other hand, for low $N_{\rm sat}$/$N_{\rm tot}$ planes, it avoids choosing subsamples of satellites that form very narrow planes but are just chance alignments, i.e., not embedded in larger planar structures.

 The quality of planes is quantified through the number of member satellites $N_{\rm sat}$ and the degree of flattening. The latter is measured through the short-to-long axis ratio of the  Tensor of Inertia \citep[ToI,][]{Metz07} concentration ellipsoid,  $c/a$
%%%(and/or the r.m.s thickness normal to the plane, $\Delta$RMS,  which are often correlated), 
provided there is a high intermediate-to-long axis ratio, $b/a$, which confirms a planar-like spatial distribution of satellites (the RMS thickness normal to the plane is correlated to $c/a$ and it does not add further information).
Quality comparisons 
between  planes
are done either considering constant $N_{\rm sat}$ (where a lower  $c/a$ means higher quality) or constant $c/a$ (where   more populated planes have higher quality). In this way, we are able to single out new high-quality planes.

This method has been applied  to the most up-to-date positional data of MW and M31 confirmed satellites
(see Tables~\ref{tab:mwdata} and \ref{tab:m31data}, and references therein).
%%%\citep[see Tables MWXXX and XXX, and also][]{McConnachie12,Pawlowski13}, 
%\textbf{updated as of October 2019,
%with $N_{\rm tot}$ = \green{46  and 34} satellites  respectively.  
In the case of the MW we consider a total of $N_{\rm tot} =$46   satellite galaxies (sample MW46). Also,
for the sake of comparison with P13's results and with numerical simulations, 
%in the case of the MW 
we have also 
studied the same satellite sample as in P13, consisting of $N_{\rm tot} =$ 27 satellites.
In the case of M31 we consider the 34 satellites within 300 kpc from M31.
Moreover, 
contemplating the possibility of M31's larger total virial mass and radius relative to the MW,
 we have extended the limiting satellite distance to 350 kpc, hence including satellites AXVI and AXXXIII in a complementary analysis.

Two  predominant, collimated  over-density regions show up 
%in each of their respective 4-galaxy-normal density plots.
in the 4-galaxy-normal density plots of both the MW and M31.
They  reveal that both their satellite systems are highly structured in planar-like configurations.
However,  they show very different patterns:
while satellites in the MW form basically one main polar structure  (either in the $N_{\rm tot}$ = 27 or 46 samples),
M31 satellites are spatially distributed along \textit{two} distinct collections of planes, inclined with respect to  the M31 disc and roughly perpendicular to each other.

We find planes of satellites with higher qualities than those previously reported  with a given $N_{\rm sat}$.
More specifically, 
we find a combination of 11 MW satellites that spatially describe a  plane with much higher quality
 than that defined 
by the 11 classical  (i.e., the most massive) satellites.
Similarly, we find combinations of $N_{\rm sat}$ =24 and 27 satellites
%Other examples have been found with $N_{\rm sat}$ =24 and 27, 
with lower $c/a$ values than the VPOS-3 or VPOSall, respectively. 
We have also found planes as thin as the VPOS-3 or VPOSall, but more populated (with $N_{\rm sat}$ = 30 and 34 satellites, respectively) and with approximately the same normal direction.  
We single out the 
plane with $N_{\rm sat}$ = 39 (MW46-1-39 in Table~\ref{table_obsThisWork}) with $c/a=0.212\pm0.002$,
as the most populated plane that thin identified so far in the Local Group; this is our most noticeable  result for the Milky Way.
Another remarkable result is that, when expressed in terms of the satellite fraction $f_{\rm sat} \equiv N_{\rm sat}/N_{\rm tot}$, results for the MW satellite sample with 46 satellites follow planar structures  found for the sample with 27 satellites (VPOS), although they differ in 21 satellite members. 
This proves how the MW's polar plane of satellites is a robust structure that endures as the census of MW satellites increases in number.
However, some words of caution are in order, as
 these results will need to be continuously revised  as new ultrafaint galaxies are discovered with upcoming deep surveys.

An analysis through the clustering of the satellite orbital poles indicates that at least  $\sim45\%$ of satellites co-orbit within these new-identified planes, with a fraction of $\sim50\%$ consistent within the errors
(this is not differentiating between co- and counter-rotating with the MW disc).
%\red{Furthermore, we study the kinematical-coherence of MW satellites in an independent way, by searching for the main axis of co-orbitation through  the '3Jorb-barycenter method' (which we apply to disc galaxy simulations in Santos-Santos et al. in prep.). Remarkably, we find that such main axis is only $\sim3^\circ$ away from the normal vector to the MW46-1-39 plane.}
  Furthemore, in order to study the kinematical-coherence of MW satellites in an independent way, we  briefly introduce the '3J$_{\rm orb}$-barycenter' method.
This method  aims at identifying orbitation axes whose normal planes contain the maximum possible number of kinematically-coherent satellites out of $N_{\rm tot}$ of them 
(the so-called '3J$_{\rm orb}$' axes). We apply this method to the samples of satellites with kinematical information drawn from MW27 and MW46, and find that in each case, a unique orbitation 
3J$_{\rm orb}$ axis stands out, with the particularities that they both point to  approximately the same direction and that this direction is only $\sim 3^\circ$ away from the normal vector to the MW46-1-39 plane. 
Clustering of orbital poles around these two 3J$_{\rm orb}$ axes of either MW27 or MW46 satellites yields results that are consistent within the error bands, and are also 
consistent with clustering around the MW46 Peak 1 direction, or the normal vector to the MW46-1-39 plane: $45\pm5$\%  satellites co-orbit around any of these axes.

%%%For M31, we present a combination of 14 satellites with much lower $c/a$   values  than those of the plane noted by \citet{Ibata13} and \citet{Conn13}  (Ibata-Conn-14 plane) with the PAndAS survey.

%%\magenta{As for M31, our most relevant result is that,}
%\textbf{More relevantly, }
As for M31,
 for the first time, 
 the second-most predominant planar structure  (Peak 2) found in M31 has been studied in detail.
 % (Peak 1 was studied in \citet{Pawlowski13}).
This peak points to a satellite planar configuration whose normal direction aligns with the line-of-sight between the Sun and  M31, and therefore is viewed nearly 
face-on (we recall that the planar configuration defined by M31 Peak 1 --containing the well-known GPoA with   $N_{\rm sat}$=19-- is viewed nearly edge-on). 
Our  analysis reveals a rich plane structure, with quality behaviour in terms of $c/a$ 
% $\Delta$RMS 
versus $N_{\rm sat}$ similar 
to that found around Peak 1, 
despite being more affected by the radial Sun - satellite distance uncertainties due to its orientation.
%Indeed, the two planar structures are well defined as revealed by the stability of their normal vector directions as $N_{\rm sat}$ increases up to around $\approx19$ members in each case.
 As quantitatively shown by similar peak strength $C_p$ values, this is  an independent planar structure to that defined by Peak 1. Indeed,  different subsamples of satellites contribute dominantly to each   structure, while there is  a $\sim$50\% of shared satellites that contribute less.

We present a combination of 14 M31 satellites from Peak 1 with much lower $c/a$   values  than those of the plane noted by \citet{Ibata13} and \citet{Conn13}  (Ibata-Conn-14 plane) with the PAndAS survey.
Also,     $c/a$ starts increasing more sharply for $N_{\rm sat}> 16$  satellites around Peak 1, and   for $N_{\rm sat}> 18$    around Peak 2.  Therefore we single out  the planes of satellites with precisely  these  $N_{\rm sat}$ values 
as they present exceptional quality. As  evidence,  the planes' normals stay stable up to $N_{\rm sat}\sim19$ for Peak 1, and $N_{\rm sat}\sim18$ for Peak 2, and then they change.
Interestingly, the plane from Peak 1 with $N_{\rm sat}$=16 
is more populated and thinner than the Ibata-Conn-14 plane. 
%%% (with $N_{\rm sat}$=14).
Moreover, the plane with $N_{\rm sat}$=18 from Peak 2
 presents 
 very similar properties to the GPoA while consisting
 of an overall different satellite sample: $c/a$ is $0.15\pm0.03$.
A first approach to its kinematics through the available line-of-sight velocities suggests a high perpendicular velocity dispersion that would soon disperse the plane. This should be confirmed however with proper motions and 3D velocity data in the future.

%%\red{ We also caution the reader about the sensibility of the  M31 Peak 2 planar structure to satellite distance determinations, which are currently very uncertain. Due to its almost face-on orientation relative to the line-of-sight from the Sun, any    changes resulting in a dispersion of order the $\Delta$RMS of the plane could imply its dissapearance. More accurate M31 satellite distances will allow to establish the robustness or not of this planar structure.}

Adding to the analysis the two further away M31 satellites placed beyond 300 kpc does not change these conclusions. Quantitatively, the $b/a$ ratios increase
(as the size of the system considered is larger)
 %and the number of satellites in planes to be singled out from Peak 1 increase in 1 member (AXVI), and in 2 members from Peak 2,
  but the rest  of the ToI parameters of Peaks 1 and 2 planes remain essentially the same.

%%%% NUEVO PARRAFO %%%
We however have to caution the reader about the sensibility of the  M31 Peak 2 planar structure to satellite distance determinations. 
Due to its almost face-on orientation relative to the line-of-sight from the Sun, any changes
in the distance moduli data used,
 resulting in a distance dispersion relative to the data assumed in this work (Table~\ref{tab:m31data})
 of the order of  a few times the $\Delta$RMS of this plane,
could possibly erase the positionally-detected Peak 2 plane.
This can be an issue, as distance moduli for a same object obtained with different methodologies can be very different 
 \citep[because the calibrations they are based on are different and not well fixed; see discussions in][]{MartinezVazquez2017,Weisz2019}.  
 %This situation could improve as the \textit{Gaia} mission goes on and parallaxes with smaller errors become available for more distant stars.
%%%%%%%%%%%%%%%%%%%%%%

Finally, 
the  richer plane structure in the MW and M31 we report in this work  was found because we allow the mass of   satellites to play no role in our search.
Indeed, through correlation tests we find 
that mass is not a satellite property that determines its 4-galaxy-normal contribution
to the main  over-density regions
 (i.e., its membership or not to the respective high-quality planes),
 either in the MW or M31 cases.
%This is expected, given that globular clusters and streams have been found to align as well with the VPOS \citep{Pawlowski12}.
%%\textbf{This could explain why}
This idea is supported by the fact that  young globular clusters and some streams
   align as well with the MW's VPOS 
   %(\citet{Pawlowski12}, but see also \citet{Riley2020}).
   \citep[see][]{Pawlowski12,Riley2020}.

We conclude that the 4GND plot method and its  quality analysis extension allow for an accurate and detailed characterization of the predominant planar spatial configurations of satellites underlying a given (observed or simulated) galactic system. 
Applied to currently available MW data, we find and characterize a unique dominant planar structure, and highlight a remarkably thin plane of satellites involving 39 over 46 satellites. Of these, $\sim45$\% are in coherent orbitation.
 Applied to M31 data, two equivalent, orthogonal  planar structures have been found, each involving $\sim19$ over 36 satellites.

\begin{table*}
\centering
\small
\caption{   ToI fitting properties  of the observed planes of satellites in the MW (classical, VPOSall, VPOS-3) and M31 (Ibata-Conn-14, GPoA):
Galactic coordinates of the normal to the plane;
uncertainty in the normal direction;
distance to the MW/M31;
root-mean-square height of the plane;
short-to-long ellipsoid axis ratio;
intermediate-to-long ellipsoid axis ratio;
number of satellites included in the plane.
 Information for the classical plane of MW satellites has been taken from
  \citet{Metz07} and \citet{Pawlowski16}.
 The information for all rest of planes has been 
  extracted from 
 \citet{Pawlowski13} (see their table 3), where distance uncertainties are considered by sampling the distances 1000 times with a Gaussian distribution around their most-likely distance.}
%%% RDT  as appears in \citet{McConnachie12}. The results presented in this table are  the mean values of averaging over all realisations.  }

\begin{tabular}{l c c c c c}
\toprule
    &   \multicolumn{3}{c}{MW} & \multicolumn{2}{c}{M31}  \\
\midrule
Name  & classical &   VPOSall & VPOS-3 & Ibata-Conn-14   & GPoA  \\
\hline
%%$\vec{n}(l,b)$ [$^\circ$]  & (-22.7, -12.7)     &(-24.4,  -3.3)  &  (-10.5,   -2.8) & (26.2,  7.8) & (25.8, 7.6)   \\
$\vec{n}(l,b)$ [$^\circ$]  & (-22.7, 12.7)     &(-24.4,  3.3)  &  (-10.5,   2.8) & (26.2,  -7.8) & (25.8, -7.6)   \\
$\Delta_{\rm sph} n$ [$^\circ$]  & --   &1.12& 0.43&  1.00&  0.79  \\
$D_{\rm MW}$ [kpc]   & 8.3   &7.9 $\pm$ 0.3 &10.4 $\pm$ 0.2   &    --   &  30.1 $\pm$ 8.8 \\
$D_{\rm M31}$ [kpc]  &  --   & 637.3 $\pm$ 13.0 &509.9 $\pm$ 10.2 &  4.1  $\pm$ 0.7  &  1.3 $\pm$ 0.6  \\
$\Delta$RMS height [kpc]&  19.6 & 29.3 $\pm$ 0.4 &19.9 $\pm$ 0.3 &    14.2  $\pm$ 0.2&         13.6 $\pm$ 0.2  \\
$c/a$             &   0.18      &0.301 $\pm$ 0.004 &0.209 $\pm$ 0.002&     0.125 $\pm$ 0.014 &        0.107 $\pm$ 0.005 \\
$b/a$           &       0.53    &0.576 $\pm$ 0.007& 0.536 $\pm$ 0.006  &  0.578 $\pm$ 0.084  &   0.615 $\pm$ 0.058  \\
$N_{\rm sat}$   &      11 & 27  &24 &  14    & 19  \\
\bottomrule
\end{tabular}
\label{table_obspl}
\end{table*}

%%%%%%%%%%%%%%%%%%%%%%%%%%%%%%%%%%%%%%%%%%%%%%%%%%n{table*}
\begin{table*}
%\centering
\scriptsize
\caption{ 
Same as Table~\ref{table_obspl} for the high quality planes singled out in this work. 
 }
\begin{tabular}{l c  c c c c c c   }
\toprule
 %%   &   \multicolumn{2}{c}{[MW-Peak-$N_{\rm sat}$]} & \multicolumn{4}{c}{[M31-Peak-$N_{\rm sat}$]}  \\
        &   \multicolumn{7}{c}{[Host-Peak-$N_{\rm sat}$]}  \\
\midrule
Name  & MW27-1-11 &  MW27-1-14 &     MW46-1-11 &  MW46-1-24 &      MW46-1-27 &  MW46-1-29 &  MW46-1-39  \\
\hline
$\vec{n}(l,b)$ [$^\circ$]  & (-9.6, -1.1)  & (-9.7, -1.1)   &    (-12.14, 1.61) & ( -11.73, 1.33) & ( -13.32, 0.39) & (-13.27, 0.60) & (-11.29, 4.24)    \\
$\Delta_{\rm sph} n$ [$^\circ$]  & 0.15  &  0.15 &0.18 & 0.16&0.16&0.17&0.26    \\
$D_{\rm MW}$ [kpc]   &15.0$\pm$0.1   &14.6$\pm$1.0   & 1.6$\pm$0.1 & 0.3$\pm$0.1 & 1.1$\pm$0.1 & 1.1$\pm$0.1 & 2.2$\pm$0.1    \\
$D_{\rm M31}$ [kpc]   &  464.5$\pm$6.3  &  465.2$\pm$6.3  &  518.2$\pm$6.9 & 511.7$\pm$6.8 & 520.3$\pm$6.9 & 520.9$\pm$6.9 & 518.6$\pm$7.0  \\
$\Delta$RMS height [kpc]  &11.6$\pm$0.1  & 10.3$\pm$0.1  & 9.8$\pm$0.1 & 9.1$\pm$0.1 &  11.8$\pm$0.1 & 12.1$\pm$0.1& 19.1$\pm$0.2  \\
$c/a$             & 0.095$\pm$0.001 & 0.096$\pm$0.001 & 0.074$\pm$0.001 & 0.094$\pm$0.001 & 0.124$\pm$0.001 & 0.131$\pm$0.01 & 0.212$\pm$0.002     \\
$b/a$           & 0.595$\pm$0.004   &0.592$\pm$0.004 &0.659$\pm$0.011 & 0.711$\pm$0.005 & 0.705$\pm$0.005 &0.702$\pm$0.005 & 0.646$\pm$0.005   \\
$N_{\rm sat}$   &      11 & 14 &11&24&27 & 29 &39  \\
\midrule
Name  &     M31\_34-1-14 & M31\_34-1-16 & M31\_34-1-23 & M31\_34-2-18   &    M31$\_$36-1-17 & M31$\_$36-1-24 & M31$\_$36-2-20 \\
\hline
$\vec{n}(l,b)$ [$^\circ$]  & (25.9, -7.2)         & (25.3, -7.4)   &  (14.4, -5.0)     &  (-68.2, 23.6)  & (25.3, -7.5) & (18.9, -5.8)   & (-60.5, 21.6)       \\
$\Delta_{\rm sph} n$ [$^\circ$]  &0.45    &0.51   & 1.40    & 3.16  & 0.30 & 0.82 & 3.14 \\
$D_{\rm MW}$ [kpc]   &   36.2$\pm$4.9 &     40.9$\pm$5.8 & 185.7$\pm$15.0   &  741.6$\pm$6.5 &  42.7$\pm$3.3 & 131.8$\pm$8.7 & 748.0$\pm$5.1  \\
$D_{\rm M31}$ [kpc]   &  1.1$\pm$0.4  & 0.6$\pm$0.5    &  2.3$\pm$1.8     & 34.9$\pm$11.2   &  0.7$\pm$0.5 &    1.8$\pm$1.2 & 38.6$\pm$12     \\
$\Delta$RMS height [kpc]  &  7.0$\pm$0.2   & 7.7$\pm$0.2   & 21.4$\pm$1.0    &21.0$\pm$4.2 &  7.2$\pm$0.2 &   22.5$\pm$0.7 &      28.7$\pm$4.8   \\
$c/a$               & 0.051$\pm$  0.002 &0.059$\pm$0.002  &0.189$\pm$0.009 &0.155$\pm$0.032  & 0.059$\pm$0.002  & 0.207$\pm$0.007 & 0.227$\pm$0.038   \\
$b/a$           & 0.553$\pm$0.046   &0.564$\pm$0.046  &0.725$\pm$0.038 &0.734$\pm$0.006 & 0.797$\pm$0.004 & 0.871$\pm$0.036 & 0.973$\pm$0.011   \\
$N_{\rm sat}$   &   14 & 16  & 23  & 18    & 17 & 24 & 20  \\
\bottomrule
\end{tabular}
\label{table_obsThisWork}
\end{table*}

\section*{Acknowledgements}
We thank the anonymous referee for suggestions that helped to improve the quality of this paper.
This work was supported through  MINECO/FEDER (Spain)  AYA2012-31101,  AYA2015-63810-P and MICIIN/FEDER (Spain) PGC2018-094975-C21 grants. 
ISS is supported by the Arthur B. McDonald Canadian Astroparticle Physics Research Institute.
This project has received funding from the European Union’s Horizon 2020 Research and Innovation Programme under the Marie Skłodowska-Curie grant agreement No 734374- LACEGAL. ISS acknowledges funding from the same Horizon 2020 grant for a secondment at the Astrophysics group of Univ. Andr\'es Bello (Santiago, Chile), and  from the Univ. Aut\'onoma de Madrid for a stay at the Leibniz Institut fur Astrophysik Potsdam (Germany). She thanks Dr. Patricia Tissera and Dr. Noam Libeskind for kindly hosting her. 
MSP thanks the DAAD for PPP grant 57512596 funded by the German Federal Ministry of Education and Research.
MSP also thanks the Klaus Tschira Stiftung gGmbH and German Scholars Organization e.V. for support via a Klaus Tschira Boost Fund.

\section*{Data availability}
The observational data for Milky Way and M31 satellite galaxies analyzed in this article comes from the following references:
\citet{McConnachie12,Conn2012,Simon2019},  
{\url{http://www.astro.uvic.ca/~alan/Nearby_Dwarf_Database_files/NearbyGalaxies.dat}}. Note that the latter is a compilation of data, and the original sources of the data are indicated therein.
The data used has been summarized for reference in this article in Tables \ref{tab:mwdata} and \ref{tab:m31data}.

%%%%%%%%%%%%%%%%%%%%%%%%%%%%%%%%%%%%%%%%%%%%%%%%%%%

\bibliographystyle{mn2e}
\bibliography{archive_planes}

\appendix

\section{MW and M31 data used in this paper} \label{sec:obsdata}

\begin{table*}
\centering
\small
\caption{ 
Data for MW confirmed satellite galaxies used in this paper, separated in the MW27 and MW46 sample. The MW46 sample includes all MW27 satellite galaxies, except for CanisMajor and BootesIII, which are marked with an asterisk.
Columns show the position in right ascension and declination, the distance modulus, the linear heliocentric distance we derive from the distance modulus, and the stellar mass (estimated assuming \citet{Woo08}'s M/L ratios according to galaxy morphological type).
 For references we refer the reader to \citet{McConnachie12}'s data compilation. 
 }
\begin{tabular}{l c  c c c c }
\toprule
GalaxyName    &  	RA/deg	   &  Dec/deg	& $(m-M)\pm$err    &  	D$_{\odot}$/kpc$^{\rm +err}_{\rm -err}$  &  	M$_{\rm star}$/M$_{\odot}$  \\
\midrule
         \multicolumn{6}{c}{MW27}  \\
\midrule
BootesI    &    210.03    &    14.50    &    19.11 $\pm$ 0.08    &    66.37 $^{ 2.49 }_{ 2.40 }$    &    3.50E+04 \\
BootesII    &    209.50    &    12.85    &    18.10 $\pm$ 0.06    &    41.69 $^{ 1.17 }_{ 1.14 }$    &    2.05E+03 \\
BootesIII*    &    209.30    &    26.80    &    18.35 $\pm$ 0.10    &    46.77 $^{ 2.20 }_{ 2.11 }$    &    2.73E+04 \\
CanesVenatici1    &    202.01    &    33.56    &    21.69 $\pm$ 0.10    &    217.77 $^{ 10.26 }_{ 9.80 }$    &    3.73E+05 \\
CanesVenaticiII    &    194.29    &    34.32    &    21.02 $\pm$ 0.06    &    159.96 $^{ 4.48 }_{ 4.36 }$    &    1.60E+04 \\
CanisMajor*    &    108.15    &    -27.67    &    14.29 $\pm$ 0.30    &    7.21 $^{ 1.07 }_{ 0.93 }$    &    7.80E+07 \\
Carina    &    100.40    &    -50.97    &    20.11 $\pm$ 0.13    &    105.20 $^{ 6.49 }_{ 6.11 }$    &    8.09E+05 \\
ComaBerenices    &    186.75    &    23.90    &    18.20 $\pm$ 0.20    &    43.65 $^{ 4.21 }_{ 3.84 }$    &    7.73E+03 \\
Draco    &    260.05    &    57.92    &    19.40 $\pm$ 0.17    &    75.86 $^{ 6.18 }_{ 5.71 }$    &    4.17E+05 \\
Fornax    &    40.00    &    -34.45    &    20.84 $\pm$ 0.18    &    147.23 $^{ 12.72 }_{ 11.71 }$    &    3.31E+07 \\
Hercules    &    247.76    &    12.79    &    20.60 $\pm$ 0.20    &    131.83 $^{ 12.72 }_{ 11.60 }$    &    2.94E+04 \\
LeoI    &    152.12    &    12.31    &    22.02 $\pm$ 0.13    &    253.51 $^{ 15.64 }_{ 14.73 }$    &    7.05E+06 \\
LeoII    &    168.37    &    22.15    &    21.84 $\pm$ 0.13    &    233.35 $^{ 14.40 }_{ 13.56 }$    &    1.08E+06 \\
LeoIV    &    173.24    &    0.53    &    20.94 $\pm$ 0.09    &    154.17 $^{ 6.52 }_{ 6.26 }$    &    1.36E+04 \\
LeoV    &    172.79    &    2.22    &    21.25 $\pm$ 0.12    &    177.83 $^{ 10.10 }_{ 9.56 }$    &    7.87E+03 \\
LMC    &    80.89    &    -69.76    &    18.52 $\pm$ 0.09    &    50.58 $^{ 2.14 }_{ 2.05 }$    &    1.06E+09 \\
PiscesII    &    344.63    &    5.95    &    21.31 $\pm$ 0.17    &    182.81 $^{ 14.89 }_{ 13.77 }$    &    6.73E+03 \\
SagittariusdSph    &    283.83    &    -30.55    &    17.10 $\pm$ 0.15    &    26.30 $^{ 1.88 }_{ 1.76 }$    &    3.44E+07 \\
Sculptor    &    15.04    &    -33.71    &    19.67 $\pm$ 0.14    &    85.90 $^{ 5.72 }_{ 5.36 }$    &    2.91E+06 \\
SegueI    &    151.77    &    16.08    &    16.80 $\pm$ 0.20    &    22.91 $^{ 2.21 }_{ 2.02 }$    &    4.53E+02 \\
SegueII    &    34.82    &    20.18    &    17.70 $\pm$ 0.10    &    34.67 $^{ 1.63 }_{ 1.56 }$    &    7.59E+02 \\
SextansI    &    153.26    &    -1.61    &    19.67 $\pm$ 0.10    &    85.90 $^{ 4.05 }_{ 3.87 }$    &    6.98E+05 \\
SMC    &    13.19    &    -72.83    &    19.03 $\pm$ 0.12    &    63.97 $^{ 3.63 }_{ 3.44 }$    &    3.23E+08 \\
UrsaMajorI    &    158.72    &    51.92    &    19.93 $\pm$ 0.10    &    96.83 $^{ 4.56 }_{ 4.36 }$    &    1.53E+04 \\
UrsaMajorII    &    132.88    &    63.13    &    17.50 $\pm$ 0.30    &    31.62 $^{ 4.69 }_{ 4.08 }$    &    6.86E+03 \\
UrsaMinor    &    227.29    &    67.22    &    19.40 $\pm$ 0.10    &    75.86 $^{ 3.58 }_{ 3.41 }$    &    5.60E+05 \\
Willman1    &    162.34    &    51.05    &    17.90 $\pm$ 0.40    &    38.02 $^{ 7.69 }_{ 6.40 }$    &    1.41E+03 \\
\midrule
         \multicolumn{6}{c}{MW46}  \\
\midrule
Antlia2    &    143.89    &    -36.77    &    20.60 $\pm$ 0.11    &    131.83 $^{ 6.85 }_{ 6.51 }$    &    5.60E+05 \\
Aquarius2    &    338.48    &    -9.33    &    20.16 $\pm$ 0.07    &    107.65 $^{ 3.53 }_{ 3.41 }$    &    7.59E+03 \\
BootesIV    &    233.69    &    43.73    &    21.60 $\pm$ 0.20    &    208.93 $^{ 20.16 }_{ 18.38 }$    &    8.87E+03 \\
Carina2    &    114.11    &    -58.00    &    17.79 $\pm$ 0.05    &    36.14 $^{ 0.84 }_{ 0.82 }$    &    8.63E+03 \\
Centaurus1    &    189.58    &    -40.90    &    20.33 $\pm$ 0.10    &    116.41 $^{ 5.49 }_{ 5.24 }$    &    2.27E+04 \\
Cetus3    &    31.33    &    -4.27    &    22.00 $\pm$ 0.20    &    251.19 $^{ 24.23 }_{ 11.31 }$    &    1.31E+03 \\
Columba1    &    82.86    &    -28.03    &    21.30 $\pm$ 0.22    &    181.97 $^{ 19.40 }_{ 17.53 }$    &    8.63E+03 \\
Crater2    &    177.31    &    -18.41    &    20.35 $\pm$ 0.02    &    117.49 $^{ 1.09 }_{ 1.08 }$    &    2.61E+05 \\
Grus2    &    331.02    &    -46.44    &    18.62 $\pm$ 0.21    &    52.97 $^{ 5.38 }_{ 4.88 }$    &    5.06E+03 \\
HorologiumI    &    43.88    &    -54.12    &    19.50 $\pm$ 0.20    &    79.43 $^{ 7.66 }_{ 6.99 }$    &    3.60E+03 \\
HorologiumII    &    49.13    &    -50.02    &    19.46 $\pm$ 0.20    &    77.98 $^{ 7.52 }_{ 6.86 }$    &    5.76E+02 \\
Hydra2    &    185.43    &    -31.99    &    20.64 $\pm$ 0.16    &    134.28 $^{ 10.27 }_{ 9.54 }$    &    9.46E+03 \\
HydrusI    &    37.39    &    -79.31    &    17.20 $\pm$ 0.04    &    27.54 $^{ 0.51 }_{ 0.50 }$    &    1.05E+04 \\
PegasusIII    &    336.09    &    5.42    &    21.56 $\pm$ 0.20    &    205.12 $^{ 19.79 }_{ 18.05 }$    &    5.76E+03 \\
PhoenixII    &    355.00    &    -54.41    &    19.60 $\pm$ 0.20    &    83.18 $^{ 8.02 }_{ 7.32 }$    &    2.86E+03 \\
Pictor2    &    101.18    &    -59.90    &    18.30 $\pm$ 0.12    &    45.71 $^{ 2.60 }_{ 3.05 }$    &    2.61E+03 \\
ReticulumII    &    53.93    &    -54.05    &    17.40 $\pm$ 0.20    &    30.20 $^{ 2.91 }_{ 2.66 }$    &    4.88E+03 \\
ReticulumIII    &    56.36    &    -60.45    &    19.81 $\pm$ 0.31    &    91.62 $^{ 14.06 }_{ 12.19 }$    &    2.88E+03 \\
Tucana2    &    342.98    &    -58.57    &    18.80 $\pm$ 0.20    &    57.54 $^{ 5.55 }_{ 5.06 }$    &    4.53E+03 \\
Tucana4    &    0.73    &    -60.85    &    18.41 $\pm$ 0.19    &    48.08 $^{ 4.40 }_{ 4.03 }$    &    3.47E+03 \\
Virgo1    &    180.04    &    0.68    &    19.80 $\pm$ 0.20    &    91.20 $^{ 8.80 }_{ 4.10 }$    &    1.85E+02 \\
\bottomrule
\end{tabular}
\label{tab:mwdata}
\end{table*}

\begin{table*}
\centering
\caption{
Data for M31 confirmed satellite galaxies used in this paper. The last two objects are AXVI and AXXXIII, satellite galaxies situated beyond 300 kpc in radial distance to M31, which we consider in the M31\_36 sample. Columns are the same as in Table~\ref{tab:mwdata}. Since for M31 satellites the distance uncertainties are important and relevant to the results presented in this work, we also make explicit in this table the distance moduli sources and the respective  methods used in each case. These are: [1] \citet{Fiorentino2010}, RR Lyrae; [2] \citet{McConnachie05}, TRGB; [3] \citet{Conn2012}, TRGB; [4] \citet{Richardson2011}, HB; [5] \citet{Yang2012}, \textit{HST}-based RR Lyrae; [6] \citet{Martin2013a}, TRGB; [7] \citet{Bell2011}, TRGB; [8] \citet{Sanna2008}, \textit{HST}-based TRGB;  [9] \citet{Martin2013b}, TRGB.
}
\begin{tabular}{l c c c c c c  }
\toprule
GalaxyName    &  	RA/deg	   &  Dec/deg	& $(m-M)\pm$err    &  	D$_{\odot}$/kpc$^{\rm +err}_{\rm -err}$  &  	M$_{\rm star}$/M$_{\odot}$ & Distance ref. \\
\midrule
         \multicolumn{7}{c}{M31\_34}  \\
\midrule
AndromedaI    &    11.42    &    38.04    &    24.36 $\pm$ 0.07    &    744.73 $^{ 24.40 }_{ 23.62 }$    &    7.59E+06    &    [2] \\
AndromedaII    &    19.12    &    33.42    &    24.07 $\pm$ 0.06    &    651.63 $^{ 18.26 }_{ 17.76 }$    &    1.46E+07    &    [2] \\
AndromedaIII    &    8.89    &    36.50    &    24.37 $\pm$ 0.07    &    748.17 $^{ 24.51 }_{ 23.73 }$    &    1.60E+06    &    [2] \\
AndromedaV    &    17.57    &    47.63    &    24.44 $\pm$ 0.08    &    772.68 $^{ 29.00 }_{ 27.95 }$    &    8.96E+05    &    [2] \\
AndromedaVI    &    357.94    &    24.58    &    24.47 $\pm$ 0.07    &    783.43 $^{ 25.67 }_{ 24.85 }$    &    5.30E+06    &    [2] \\
AndromedaVII    &    351.63    &    50.68    &    24.41 $\pm$ 0.10    &    762.08 $^{ 35.92 }_{ 34.30 }$    &    2.63E+07    &    [2] \\
AndromedaIX    &    13.22    &    43.20    &    24.42 $\pm$ 0.07    &    765.60 $^{ 25.08 }_{ 24.29 }$    &    2.42E+05    &    [2] \\
AndromedaX    &    16.64    &    44.80    &    24.13 $\pm$ 0.08    &    669.88 $^{ 25.14 }_{ 38.93 }$    &    1.41E+05    &    [3] \\
AndromedaXI    &    11.58    &    33.80    &    24.33 $\pm$ 0.05    &    734.51 $^{ 17.11 }_{ 16.72 }$    &    7.38E+04    &    [5] \\
AndromedaXII    &    11.86    &    34.37    &    24.84 $\pm$ 0.09    &    928.97 $^{ 39.31 }_{ 134.64 }$    &    5.65E+04    &    [3] \\
AndromedaXIII    &    12.96    &    33.00    &    24.62 $\pm$ 0.05    &    839.46 $^{ 19.55 }_{ 19.11 }$    &    5.55E+04    &    [5] \\
AndromedaXIV    &    12.90    &    29.70    &    24.50 $\pm$ 0.06    &    794.33 $^{ 22.25 }_{ 180.57 }$    &    3.77E+05    &    [3] \\
AndromedaXV    &    18.58    &    38.12    &    23.98 $\pm$ 0.26    &    625.17 $^{ 79.52 }_{ 33.61 }$    &    7.73E+05    &    [3] \\
AndromedaXVII    &    9.28    &    44.32    &    24.31 $\pm$ 0.11    &    727.78 $^{ 37.82 }_{ 26.32 }$    &    3.47E+05    &    [3] \\
AndromedaXIX    &    4.88    &    35.04    &    24.57 $\pm$ 0.08    &    820.35 $^{ 30.79 }_{ 147.37 }$    &    5.30E+05    &    [3] \\
AndromedaXX    &    1.88    &    35.13    &    24.35 $\pm$ 0.12    &    741.31 $^{ 42.12 }_{ 52.66 }$    &    3.95E+04    &    [3] \\
AndromedaXXI    &    358.70    &    42.47    &    24.59 $\pm$ 0.06    &    827.94 $^{ 23.20 }_{ 26.26 }$    &    1.13E+06    &    [3] \\
AndromedaXXII    &    21.92    &    28.09    &    24.82 $\pm$ 0.07    &    920.45 $^{ 30.16 }_{ 140.62 }$    &    7.31E+04    &    [3] \\
AndromedaXXIII    &    22.34    &    38.72    &    24.43 $\pm$ 0.13    &    769.13 $^{ 47.45 }_{ 44.69 }$    &    1.69E+06    &    [4] \\
AndromedaXXIV    &    19.62    &    46.37    &    23.89 $\pm$ 0.12    &    599.79 $^{ 34.08 }_{ 32.25 }$    &    1.49E+05    &    [4] \\
AndromedaXXV    &    7.54    &    46.85    &    24.55 $\pm$ 0.12    &    812.83 $^{ 46.18 }_{ 43.70 }$    &    1.09E+06    &    [4] \\
AndromedaXXVI    &    5.94    &    47.92    &    24.41 $\pm$ 0.12    &    762.08 $^{ 43.30 }_{ 40.97 }$    &    9.55E+04    &    [4] \\
AndromedaXXVII    &    9.36    &    45.39    &    24.59 $\pm$ 0.12    &    827.94 $^{ 47.04 }_{ 44.51 }$    &    1.96E+05    &    [4] \\
AndromedaXXIX    &    359.73    &    30.76    &    24.32 $\pm$ 0.22    &    731.14 $^{ 77.96 }_{ 70.45 }$    &    2.91E+05    &    [7] \\
AndromedaXXX    &    9.15    &    49.65    &    24.17 $\pm$ 0.10    &    682.34 $^{ 32.16 }_{ 77.00 }$    &    2.11E+05    &    [3] \\
AndromedaXXXI    &    344.57    &    41.29    &    24.40 $\pm$ 0.12    &    758.58 $^{ 43.10 }_{ 40.78 }$    &    6.55E+06    &    [6] \\
AndromedaXXXII    &    9.00    &    51.56    &    24.45 $\pm$ 0.14    &    776.25 $^{ 51.70 }_{ 48.47 }$    &    1.09E+07    &    [6] \\
IC10    &    5.07    &    59.30    &    24.50 $\pm$ 0.12    &    794.33 $^{ 45.13 }_{ 42.71 }$    &    1.37E+08    &    [8] \\
LGS3    &    15.98    &    21.89    &    24.43 $\pm$ 0.07    &    769.13 $^{ 25.20 }_{ 24.40 }$    &    1.54E+06    &    [2] \\
M32    &    10.67    &    40.87    &    24.53 $\pm$ 0.21    &    805.38 $^{ 81.78 }_{ 74.24 }$    &    2.23E+08    &    [1] \\
NGC147    &    8.30    &    48.51    &    24.15 $\pm$ 0.09    &    676.08 $^{ 28.61 }_{ 27.45 }$    &    9.91E+07    &    [2] \\
NGC185    &    9.74    &    48.34    &    23.95 $\pm$ 0.09    &    616.60 $^{ 26.09 }_{ 25.03 }$    &    1.09E+08    &    [2] \\
NGC205    &    10.09    &    41.69    &    24.58 $\pm$ 0.07    &    824.14 $^{ 27.00 }_{ 26.14 }$    &    5.35E+08   &    [2] \\
Triangulum    &    23.46    &    30.66    &    24.54 $\pm$ 0.06    &    809.10 $^{ 22.67 }_{ 22.05 }$    &    2.06E+09    &    [2] \\
\midrule
         \multicolumn{7}{c}{M31\_36}  \\
\midrule
AndromedaXVI    &    14.87    &    32.38    &    23.39 $\pm$ 0.19    &    476.43 $^{ 43.57 }_{ 29.75 }$    &    5.40E+05    &    [3] \\
AndromedaXXXIII    &    45.35    &    40.99    &    24.49 $\pm$ 0.18    &    790.68 $^{ 68.33 }_{ 62.90 }$    &    1.96E+06    &    [9] \\
\bottomrule
\end{tabular}
\label{tab:m31data}
\end{table*}

\label{lastpage}
 
\end{document}